\documentclass[twocolumn]{aastex62}
\usepackage{float}
\usepackage{amssymb,amsmath,natbib,gensymb}
\usepackage{color}
\usepackage{ulem}
\usepackage{latexsym}
\usepackage{wasysym}   % Waldis symbol fonts containing \astrosun

\shorttitle{ASASSN-15pz}
\shortauthors{Ping Chen et al.}

\begin{document}

\title{ASASSN-15pz: Revealing Significant Photometric Diversity among 2009dc-like, Peculiar SNe Ia \footnote{This paper includes data gathered with the 6.5 meter Magellan Telescopes located at Las Campanas Observatory, Chile}}
\author[0000-0003-0853-6427]{Ping Chen}\affil{Kavli Institute for Astronomy and Astrophysics, Peking University, Yi He Yuan Road 5, Hai Dian District, Beijing 100871, China.}\affil{Department of Astronomy, School of Physics, Peking University, Yi He Yuan Road 5, Hai Dian District, Beijing 100871, China}
\author[0000-0002-1027-0990]{Subo Dong},\affil{Kavli Institute for Astronomy and Astrophysics, Peking University, Yi He Yuan Road 5, Hai Dian District, Beijing 100871, China.}
\author{Boaz Katz}\affil{Department of Particle Physics and Astrophysics, Weizmann Institute of Science, Rehovot 76100, Israel}
\author{C. S. Kochanek} \affil{Department of Astronomy, The Ohio State University, 140 West 18th Avenue, Columbus, OH 43210, USA}\affil{Center for Cosmology and AstroParticle Physics, The Ohio State University, 191 W. Woodruff Ave., Columbus, OH 43210, USA}
\author{Juna A. Kollmeier} \affil{Observatories of the Carnegie Institution for Science, 813 Santa Barbara Street, Pasadena, CA 91101, USA}
\author[0000-0002-9770-3508]{K. Maguire}\affil{Astrophysics Research Centre, School of Mathematics and Physics, Queen's University Belfast, Belfast BT7 1NN, UK}\affil{School of Physics, Trinity College Dublin, Dublin 2, Ireland}
\author{M. M. Phillips}\affil{Las Campanas Observatory, Carnegie Observatories, Casilla 601, La Serena, Chile}
\author{J. L. Prieto}\affil{N\'ucleo de Astronom\'ia, Facultad de Ingenier\'ia y Ciencias, Universidad Diego Portales, Ej\'ercito 441, Santiago, Chile}\affil{Millennium Institute of Astrophysics, Santiago, Chile}
\author{B. J. Shappee}\affil{Institute for Astronomy, University of Hawaii, 2680 Woodlawn Drive, Honolulu, HI 96822, USA}
\author[0000-0002-5571-1833]{M. D. Stritzinger} \affil{Department of Physics and Astronomy, Aarhus University, Ny Munkegade 120, DK-8000 Aarhus C, Denmark }
\author{Subhash Bose},\affil{Kavli Institute for Astronomy and Astrophysics, Peking University, Yi He Yuan Road 5, Hai Dian District, Beijing 100871, China.}
\author[0000-0001-6272-5507]{Peter J. Brown}\affil{Department of Physics and Astronomy,Texas A$\&$M University, 4242 TAMU, College Station, TX 77843, USA }\affil{George P. and Cynthia Woods Mitchell Institute for Fundamental Physics $\&$ Astronomy}
\author[0000-0001-9206-3460]{T.~W.-S.~Holoien}\affiliation{The Observatories of the Carnegie Institution for Science, 813 Santa Barbara St., Pasadena, CA 91101, USA}
\author{L. Galbany}\affil{PITT PACC, Department of Physics and Astronomy, University of Pittsburgh, Pittsburgh, PA 15260, USA}
\author{Peter A. Milne}\affil{University of Arizona, Steward Observatory, 933 North Cherry Avenue, Tucson, AZ 85719, USA}
\author{Nidia Morrell}\affil{Las Campanas Observatory, Carnegie Observatories, Casilla 601, La Serena, Chile}
\author{Anthony L. Piro}\affil{Observatories of the Carnegie Institution for Science, 813 Santa Barbara Street, Pasadena, CA 91101, USA}
\author{K. Z. Stanek} \affil{Department of Astronomy, The Ohio State University, 140 West 18th Avenue, Columbus, OH 43210, USA}\affil{Center for Cosmology and AstroParticle Physics, The Ohio State University, 191 W. Woodruff Ave., Columbus, OH 43210, USA}
\author{Todd A. Thompson}\affil{Department of Astronomy, The Ohio State University, 140 West 18th Avenue, Columbus, OH 43210, USA}\affil{Center for Cosmology and AstroParticle Physics, The Ohio State University, 191 W. Woodruff Ave., Columbus, OH 43210, USA}\affil{Institute for Advanced Study, 1 Einstein Drive, Princeton, NJ 08540, USA}
\author{D. R. Young}\affil{Astrophysics Research Centre, School of Mathematics and Physics, Queens University Belfast, Belfast BT7 1NN, UK}
\correspondingauthor{Subo Dong}
\email{dongsubo@pku.edu.cn}

\begin{abstract}

We report comprehensive multi-wavelength observations of a peculiar Type Ia-like supernova (``SN Ia-pec'') ASASSN-15pz. ASASSN-15pz is a spectroscopic ``twin'' of SN 2009dc, a so-called ``Super-Chandrasekhar-mass'' SN, throughout its evolution, but it has a peak luminosity $M_{B, \rm peak}  = -19.69\pm0.12\,{\rm mag}$ that is {$\approx 0.6\,{\rm mag}$} dimmer and comparable to the SN~1991T sub-class of SNe~Ia at the luminous end of the normal width-luminosity relation. The synthesized $^{56}$Ni mass of $M_{{^{56}}{\rm Ni}} = {1.13} \pm 0.14 M_\odot$ is also substantially less than that found for several 2009dc-like SNe. Previous well-studied 2009dc-like SNe have generally suffered from large and uncertain amounts of host-galaxy extinction, which is negligible  for ASASSN-15pz. Based on the color of ASASSN-15pz, we estimate a host extinction for SN 2009dc of $E(B-V)_{\rm host}={0.12}\,{\rm mag}$ and confirm its high luminosity ($M_{B, \rm peak}[{\rm 2009dc}] \approx {-20.3} \,{\rm mag}$). The 2009dc-like SN population, which represents $\sim 1\%$ of SNe~Ia, exhibits a range of peak luminosities, and do not fit onto the tight width-luminosity relation. Their optical light curves also show significant diversity of late-time ($\gtrsim 50$\,days) decline rates. The nebular-phase spectra provide powerful diagnostics to identify the 2009dc-like events as a distinct class of SNe Ia. We suggest referring to these sources using the phenomenology-based ``2009dc-like SN Ia-pec'' instead of ``Super-Chandrasekhar SN Ia,''  which is based on an uncertain theoretical interpretation.

\end{abstract}


\section{INTRODUCTION}
\begin{figure}
\centerline{\includegraphics[width=10cm,height=5.5cm]{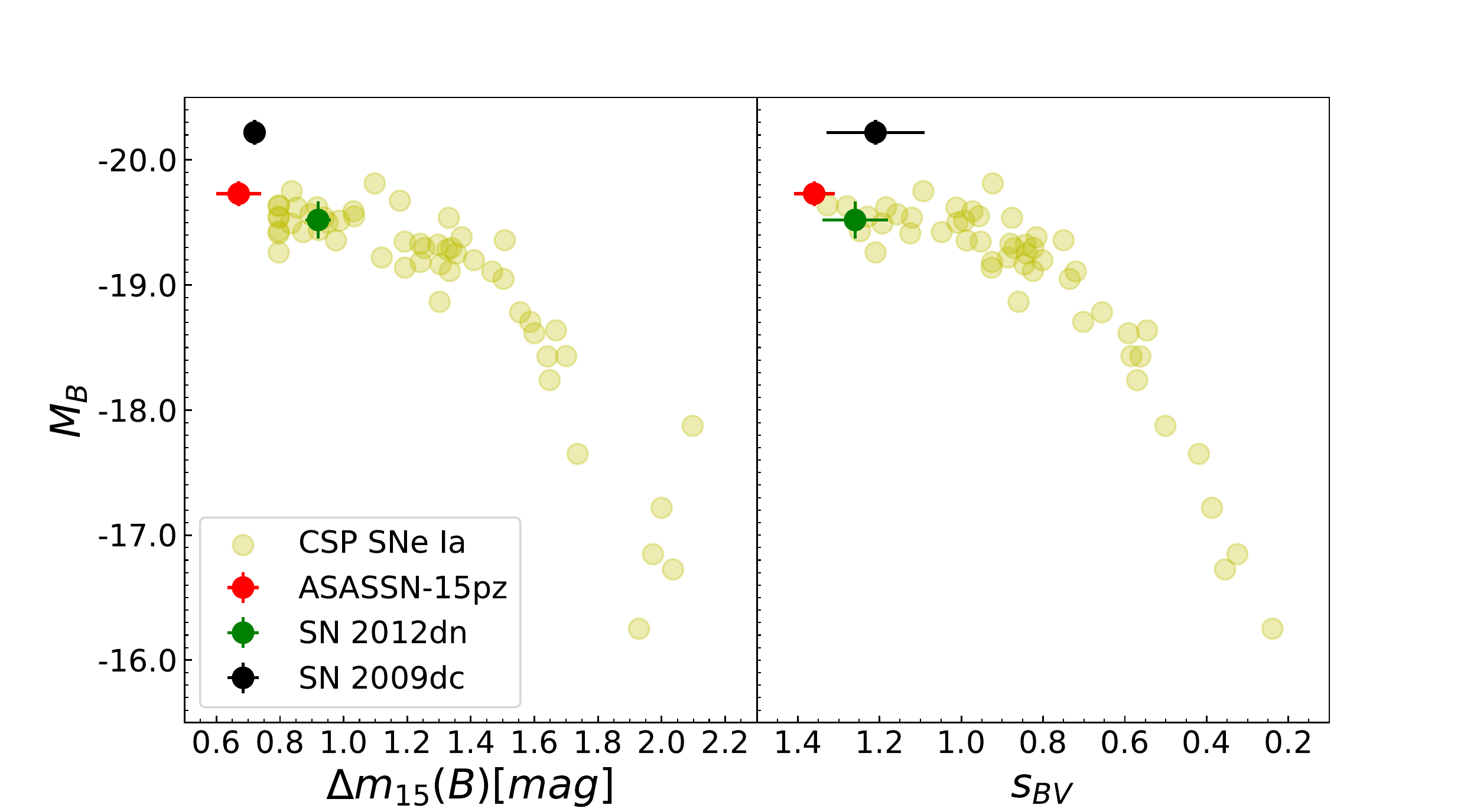}}
\caption{Width vs. Luminosity  (decline-rate) relation \citep{Phillips1993} populated with a subset of CSP-I SNe Ia from  \cite{Burns2014}. The relation is parameterized by $\Delta m_{15}$ (left) and $s_{BV}$ (right). The absolute $B$-band magnitudes are extinction-corrected
as described by  \cite{Burns2014}, and distances are computed using a Hubble constant of $H_0$ = 73\,{\rm km s}$^{-1}$\,{\rm Mpc}$^{-1}$ for SNe~Ia with redshift $z > 0.01$ and redshift-independent distances from the literature for objects with $z < 0.01$.}
\label{fig:MB_m15_sbv}
\end{figure}

The empirical correlations between peak luminosity and light-curve width (or the post-peak decline rate; see the yellow filled circles in the left panel of Fig.~\ref{fig:MB_m15_sbv}) established by \citet{Phillips1993} make Type Ia supernovae (SNe~Ia) important standardizable candles for cosmology. Yet, the progenitor systems and explosion mechanism of SNe~Ia are still under intense debate \citep[e.g., see the reviews by][]{Maoz2014,Wang2018}. The width-luminosity relation \citep{Pskovskii1977, Phillips1993, Phillips2005} may also hold important clues to understanding the physics of the SNe~Ia population \citep[e.g.,][]{Wygoda2019a, Wygoda2019b}, and these relations appear to connect the dimmest 1991bg-like events \citep{Filippenko1992_91bg,Leibundgut1993,Turatto1996}, the most luminous 1991T-like events \citep{Filippenko1992_91T, Jeffery1992,Ruiz1992}, and the ``normal'' SNe~Ia used for cosmology into a continuous distribution (e.g., see Figure ~4 of \citealt{Burns2018} and the yellow filled circles in the right panel of Fig.~\ref{fig:MB_m15_sbv}). Evidence for the continuity of the SN~Ia population includes other light-curve properties \citep[e.g.][]{Phillips2012, Burns2014}, early-phase spectroscopic properties \citep[e.g.,][]{Nugent95, Branch09}, and nebular-phase spectroscopic properties \citep[e.g.,][]{Mazzali1998,Kushnir2013,Dong2018}. 

In recent years, wide-field sky surveys have discovered an increasing number of peculiar and luminous SNe~Ia-like events. These objects are spectroscopically similar to SNe~Ia, especially the 1991T-like sub-class, near maximum light, but they often exhibit spectroscopic peculiarities and also usually deviate from the Phillips relation \citep[see, e.g.,][]{Taubenberger2017}. In particular, there are SNe~Ia-like objects with peak absolute magnitudes  of $M_V^{peak}\simeq -20$, that are greater than those of the most luminous SNe~Ia (i.e., the 1991T-like sub-class) on the Phillips relation, leading to speculation that their progenitor masses may be super-Chandrasekhar. \cite{Howell2006} found the first such event, SN~2003fg, and a small number of similarly luminous SNe~Ia-like events (referred to in the following text as luminous peculiar SNe~Ia or ``luminous SNe~Ia-pec'' in short) have been discovered later: SN~2006gz \citep{Hicken2007}, SN~2007if \citep{Scalzo2010ec} and SN~2009dc \citep{Yamanaka2009, Silverman2011, Taubenberger2011}. Their spectra have substantial differences from SNe~Ia on the Phillips relation. In general, they appear to be hotter and show evidence of unburned carbon in their early-phase spectra and weak or even absent [\ion{Fe}{3}] emission in their nebular spectra. The luminosity estimates for these objects, including SN 2009dc, have suffered from uncertainties in the host-galaxy extinction. In the case of SNe~Ia on the Phillips relation, there is a remarkable uniformity to the intrinsic ($B-V$) color between 30 and 90 days (i.e., the Lira relation; \citealt{Phillips1999}) for objects with negligible host extinction, allowing accurate estimates for host-reddened SNe~Ia. However, the luminous SNe~Ia-pec found to date all suffer from significant host-reddening and their intrinsic colors remain unclear. 

In particular, the host extinction of the best studied example, SN 2009dc, is uncertain. Its existence is clear due to the presence of a \ion{Na}{1}~D absorption line with an equivalent width of EW(\ion{Na}{1}~D) $\approx 1.0$\,\AA\,\citep{Silverman2011,Taubenberger2011}. \citet{Taubenberger2011} adopted an extinction of $E(B-V)_{\rm host}=0.10\pm0.07$ mag. They noted that the post-peak ($B-V$) curve of SN~2009dc does not follow the Lira relation and is redder by $\sim 0.3$\,mag at 90\,d. Considering various empirical relations between EW(\ion{Na}{1}~D) and dust extinction, \cite{Silverman2011} found a range of $0.1<E(B-V)_{\rm host} <0.3$ mag that translates into a significant uncertainty of $\sim 0.6\,$mag in the peak luminosity. Another luminous SN Ia-pec, SN 2006gz, also suffers from the uncertainties in its host-galaxy extinction. If the Lira relation is adopted, the host extinction is $E(B-V)_{\rm host}=0.18$ mag, implying a peak absolute magnitude of $M_V^{\rm peak}=-19.74\pm0.16$ \citep{Hicken2007}.  SN 2012dn was found to be a spectroscopic ``clone'' of SN 2006gz \citep{Chakradhari2014}, while its estimated peak luminosities of $M_B^{\rm peak}=-19.52\pm0.15\,{\rm mag}$ and $M_V^{\rm peak}=-19.42\pm0.15\,{\rm mag}$ were similar to an average SN Ia but substantially dimmer than previously known 2009dc-like SNe. SN 2012dn was also subject to large uncertainty in host extinction estimates. 

Here we report the discovery and observations of ASASSN-15pz, which is nearly identical to SN~2009dc spectroscopically, but has some significantly different photometric properties including its peak luminosity and late-time light-curve decline rates.  Our UV, optical, and near-infrared (NIR) photometry data and visual-wavelength spectra of ASASSN-15pz are described in \S~\ref{data}. In \S~\ref{photometric_results}, we show that ASASSN-15pz is essentially free from host-galaxy extinction, allowing the luminosity and intrinsic color evolution of ASASSN-15pz to be accurately determined and compared to other SNe. Our spectroscopic data span from before the peak into the nebular phase, and we show in \S~\ref{spectra} that ASASSN-15pz is a spectroscopic ``twin'' of SN~2009dc (see \S~\ref{spectra}). We discuss the implications of our work for understanding this class of SNe~Ia in \S~\ref{summary_discussion}.

\section{Observations and Data Reduction}
\label{data}
\begin{figure}
\centerline{\includegraphics[width=8cm,height=8cm]{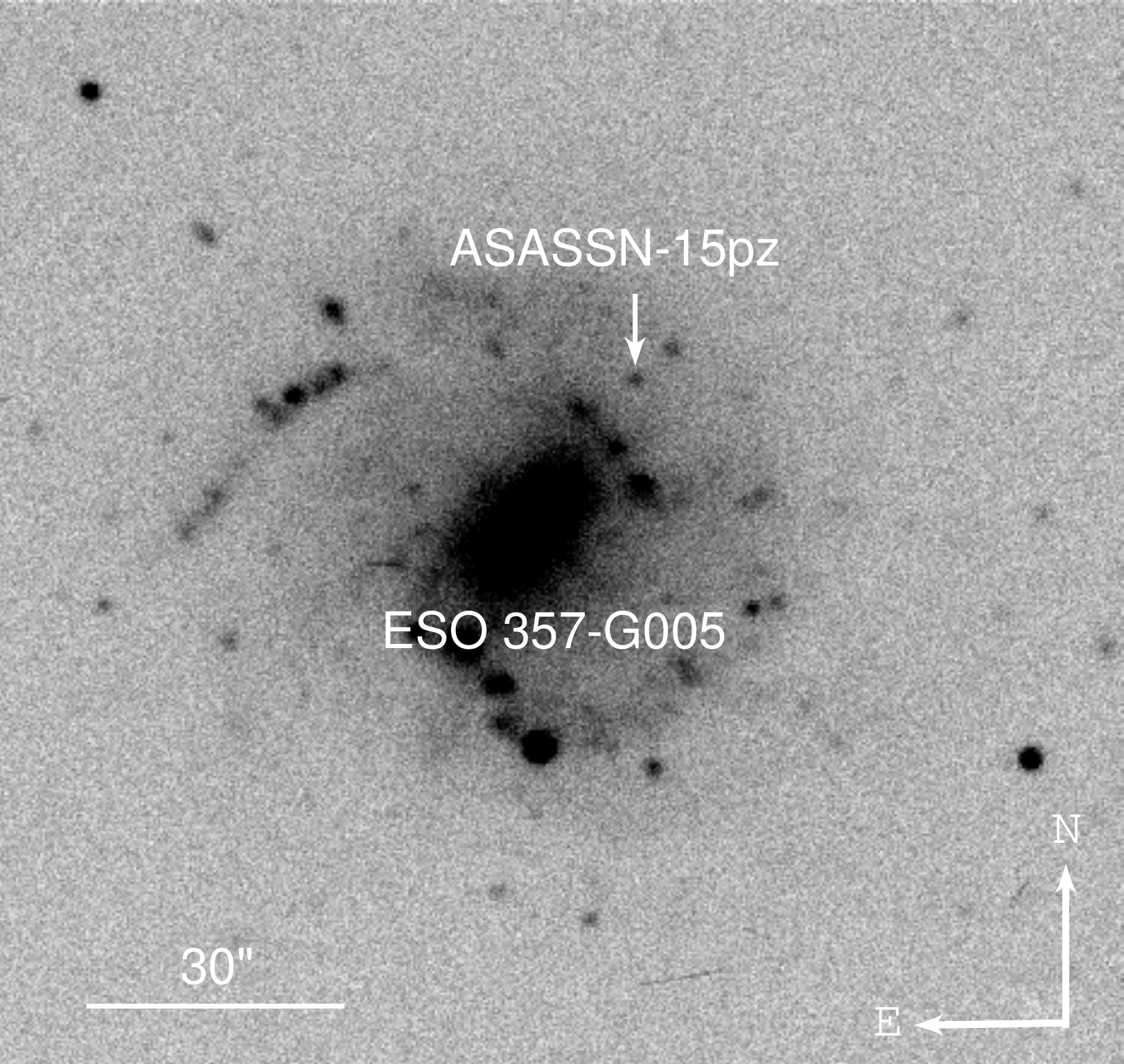}}
\caption{SDSS $r$-band image of ASASSN-15pz in the host-galaxy ESO 357-G005. The image was taken with Clay/LDSS3C on UT 2016 June 7.40 (JD = 2457546.90).}
\label{host_15pz}
\end{figure}

 ASASSN-15pz (R.A. = $03^\textnormal{h}08^\textnormal{m}48^\textnormal{s}.443$, decl. = $-35^\textnormal{d}13^\textnormal{m}50^\textnormal{s}.89$) was discovered by the All-Sky Automated Survey for SuperNovae (ASAS-SN; \citealt{Shappee2014}) on UT 2015 September 27.16 (JD = 2457292.66) at  an apparent $V$-band magnitude of $\sim$ 16.4 mag  \citep{atel_discovery}. The source is approximately 16\farcs34 North and 14\farcs45 West of the nearby SB(rs)d galaxy ESO 357-G005 (6dF J0308495-351409) at z $=$ 0.014837 \citep{Jones2009}. An image of ASASSN-15pz is shown in Fig.~\ref{host_15pz}, and its projected distance is $6.5$\,kpc from the center of the host galaxy. Correcting for the infall velocity of the Local Group toward the Virgo Cluster leads to a luminosity distance of $d_L=58.8\pm3.3$\,Mpc (distance modulus $\mu=33.85\pm0.12$\,mag) for $H_0 = 73\,{\rm km s}^{-1}\,{\rm Mpc}^{-1}$ \citep{Burns2018}, $\Omega_m=0.27$ and $\Omega_{\Lambda}=0.73$. For consistency, the distance moduli for all objects in Table~\ref{09dc_paras} are computed in the same way as for ASASSN-15pz. We include an uncertainty of $\sigma_{cz}=250~{\rm km\,s^{-1}}$ \citep[see, e.g.][]{Burns2018} to account for peculiar velocities. 
 
ASASSN-15pz was classified as a SN~Ia by an optical spectrum taken with Nordic Optical Telescope (NOT) on UT 2015 September 30.16 \citep{atel_classification}. We performed optical  follow-up imaging observations starting from its discovery as part of systematic efforts to follow up a large, complete and volume-limited $(z<0.02)$ complete sample of SNe~Ia (Chen, P. et al. 2019, in prep). It peaked in $B$ band at $14.2$\,mag on UT 2015 October 11.72 (JD 2457307.22), which is used as the phase reference epoch throughout the text. The presence of a prominent \ion{C}{2} absorption feature redward of the \ion{Si}{2} $\lambda$6355 feature, similar to the luminous SNe Ia-pec SN~2006gz and SN~2009dc, motivated us to follow this event extensively. We initiated   \textit{Neil Gehrels Swift Observatory} (\textit{Swift}; \citealt{Gehrels2004}) Ultraviolet Optical Telescope (UVOT) observation around its peak and NIR photometric observations around 10 days after $B$-band maximum. A total of eight optical spectra were obtained for ASASSN-15pz during the follow-up campaign, spanning from before the peak and into the nebular phase. We summarize our photometric and spectroscopic observations in Appendices A and B, with the photometric results reported in Tables~\ref{tab:LCOGT_BVri_mags_02},~\ref{tab:NIR_mags} and~\ref{tab:Swift_mags} and the spectroscopic observing logs given in Table~\ref{tab:speclog}.

\begin{figure}
\centerline{\includegraphics[width=10cm,height=10cm]{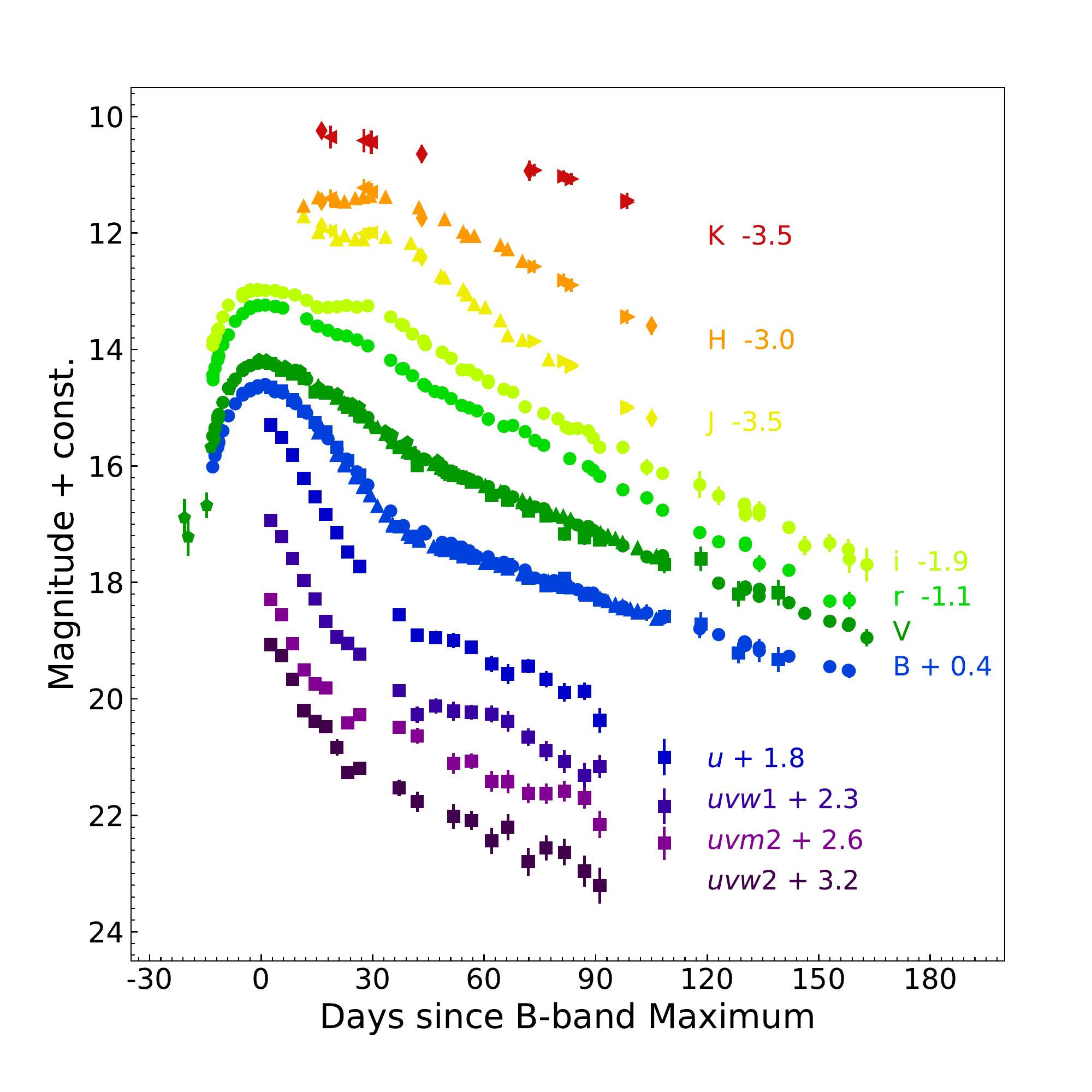}}
\caption{Multi-band light curves of ASASSN-15pz. This includes data obtained with \textit{Swift} (squares), LCOGT (circles),  SMARTS (upward-facing triangles), NOT (diamonds), and UKIRT (rightward-facing triangles for WFCAM and left-facing triangles for UFTI).}
\label{lcs_15pz}
\end{figure}

\section{Photometric evolution of ASASSN-15pz}
\label{photometric_results}
The UV, optical, and NIR light curves of ASASSN-15pz are shown in Fig.~\ref{lcs_15pz}. Our extensive photometric coverage makes it one of the best observed luminous SNe~Ia-pec. 

\subsection{Optical and NIR Light Curves}
\label{optical_lcs}

A polynomial fit to the $B$-band light curve around maximum brightness yields $B_{max} = 14.23 \pm 0.02$\,mag on JD 2457307.2$\pm$0.8 (UT 2015 Oct 11.7). Our multi-band optical photometry observations span the phases from $-$13 days to $+$163.0 days. The $V$-band light curve peaked at $14.23 \pm 0.01$\,mag on JD $2457307.2 \pm 0.7$. We derive an observed $\Delta m_{15}(B) = 0.67 \pm 0.07$ mag, which makes ASASSN-15pz one of the slowest-declining SNe~Ia known. The decline rate is similar to other luminous SNe~Ia-pec such as SN~2006gz ($\Delta m_{15}(B) = 0.69\pm0.04$ mag; \citealt{Hicken2007}),  and  
SN~2009dc ($\Delta m_{15}(B) = 0.72\pm 0.03$ mag; \citealt{Taubenberger2011}), while it is slower than 
SN~2012dn ($0.92\pm0.04$ mag; \citealt{Chakradhari2014}). Other key photometric parameters derived from the optical light curves of ASASSN-15pz are tabulated in Table~\ref{lc_paremeters}. 

\begin{table*}
\caption{Photometric parameters of ASASSN-15pz}
\begin{center}
\small
\begin{tabular}{ccccccc}
	\hline
	Band & $JD^{\rm peak}$ & $m^{\rm peak}_{\lambda}$ & $M^{\rm peak}_{\lambda}$ & $\Delta m_{15}(\lambda)$ & Decline rate$^a$ & Color at $B_{max}$ \\ 
	\hline
	$B$ & 2457307.22 $\pm$ 0.77& 14.23$\pm$0.02 & $-$19.69$\pm$0.12 & 0.67 $\pm$ 0.08&2.265& ... \\
	$V$ & 2457307.15 $\pm$ 0.71& 14.23$\pm$0.01 & $-$19.67$\pm$0.12 & 0.38 $\pm$ 0.04&2.712& $(B-V)_0=0.00\pm0.02$\\
	$r$  & 2457308.25 $\pm$ 1.25& 14.36$\pm$0.02 & $-$19.53$\pm$0.12 & 0.39 $\pm$ 0.05&3.277& $(V-r)_0=-0.13\pm0.03$\\ 
	$i$  & 2457307.65 $\pm$ 1.88& 14.86 $\pm$0.02& $-$19.02$\pm$0.12 & 0.28 $\pm$ 0.05&3.432& $(r-i)_0=-0.50\pm0.03$\\
	\hline
	\hline
\end{tabular}
\end{center}
$^a$Decline rate during 50--100 days after $B$-band maximum, in units of mag $(100d)^{-1}$.
\label{lc_paremeters}
\end{table*}

\begin{figure}
\centerline{\includegraphics[width=10cm,height=10cm]{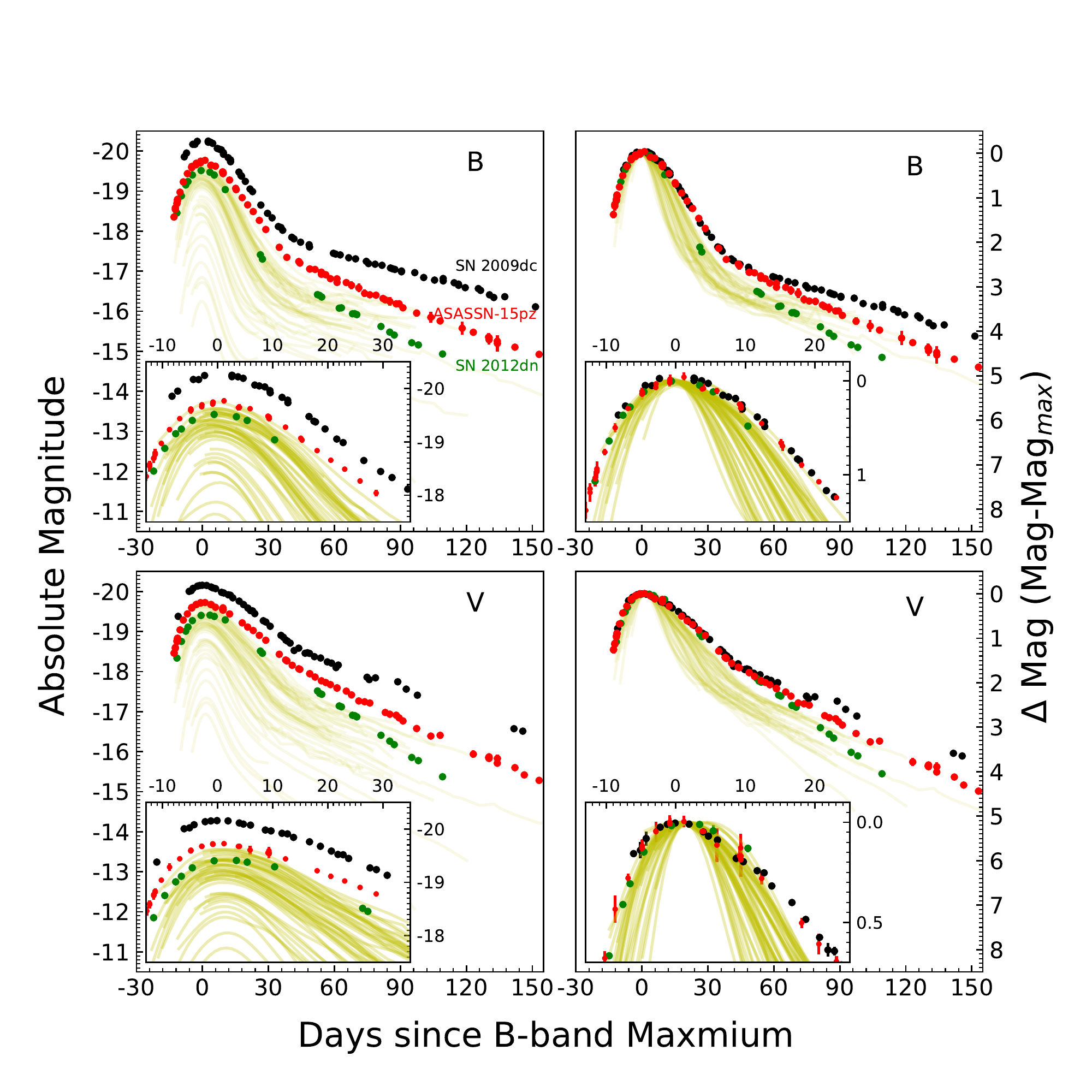}}
\caption{$B$- (top) and $V$-band (bottom) light curves of ASASSN-15pz (red) compared to those of SN~2009dc (black), SN~2012dn (green) and  SNe~Ia from CSP-I (yellow). The left two panels show absolute magnitude light curves after correcting for Galactic and host-galaxy extinctions, while in the right two panels, all the light curves have been shifted to match at their peak magnitudes and to the time of $B$-band maximum.}
\label{BV_lcs}
\end{figure}

In Fig.~\ref{BV_lcs}, we compare the $B$- and $V$-band light curves of ASASSN-15pz (red filled circles) to those of SN~2009dc (black filled circles), SN 2012dn (green filled circles), and the ``normal'' SNe Ia (yellow lines) from the \textit{Carnegie Supernova Project-I} (CSP-I; \citealt{Burns2014}). In the left two panels, the absolute-magnitude light curves are corrected for both the Milky Way (MW) and host-galaxy extinctions. For SN~2009dc, SN~2012dn, and ASASSN-15pz, the adopted reddening parameters and distance modulus are listed in Table \ref{09dc_paras}, and we assume $R_V$=3.1. We discuss how the reddening parameters are derived in \S~\ref{as15pz_extinction}. For the CSP-I sample, we used the extinction corrections from  \cite{Burns2014}. ASASSN-15pz is comparable to the most luminous SNe~Ia in the CSP-I sample, while it is significantly dimmer than SN~2009dc. SN~2012dn is $\sim 0.2$\,mag dimmer than ASASSN-15pz and is consistent with the typical 1991T/1999aa sub-class of SNe~Ia. In the right two panels, the $B$- and $V$-band light curves are normalized to the fluxes at peak. The insets highlight the light-curve shapes near maximum, and ASASSN-15pz and SN 2009dc have the broadest light curves. They also stand out in the left panel of Fig.~\ref{fig:MB_m15_sbv} when using $\Delta{m15(B)}$ to measure post-peak light-curve decline rate. 
ASASSN-15pz and SN 2009dc have nearly identical $B$- and $V$-band light-curve shapes in the early phases ($\lesssim50$ days). 
Between 50 and 150 days, ASASSN-15pz declines in the $B$ band ($V$ band) at a rate of 0.023 mag day$^{-1}$ (0.027 mag day$^{-1}$), which is substantially faster than SN~2009dc (0.015 mag day$^{-1}$ in $B$ and 0.019 mag day$^{-1}$ in $V$; \citealt{Silverman2011}). The late-time decline rate of ASASSN-15pz is comparable to the fastest declining SNe in the CSP-I sample. At late times, SN 2012dn declines even faster than ASASSN-15pz and all SNe~Ia in the CSP-I sample. \citet{Chakradhari2014} argued that the fast late-time decline of SN~2012dn is due to dust formation.  

\begin{figure}
\centerline{\includegraphics[width=10cm,height=10cm]{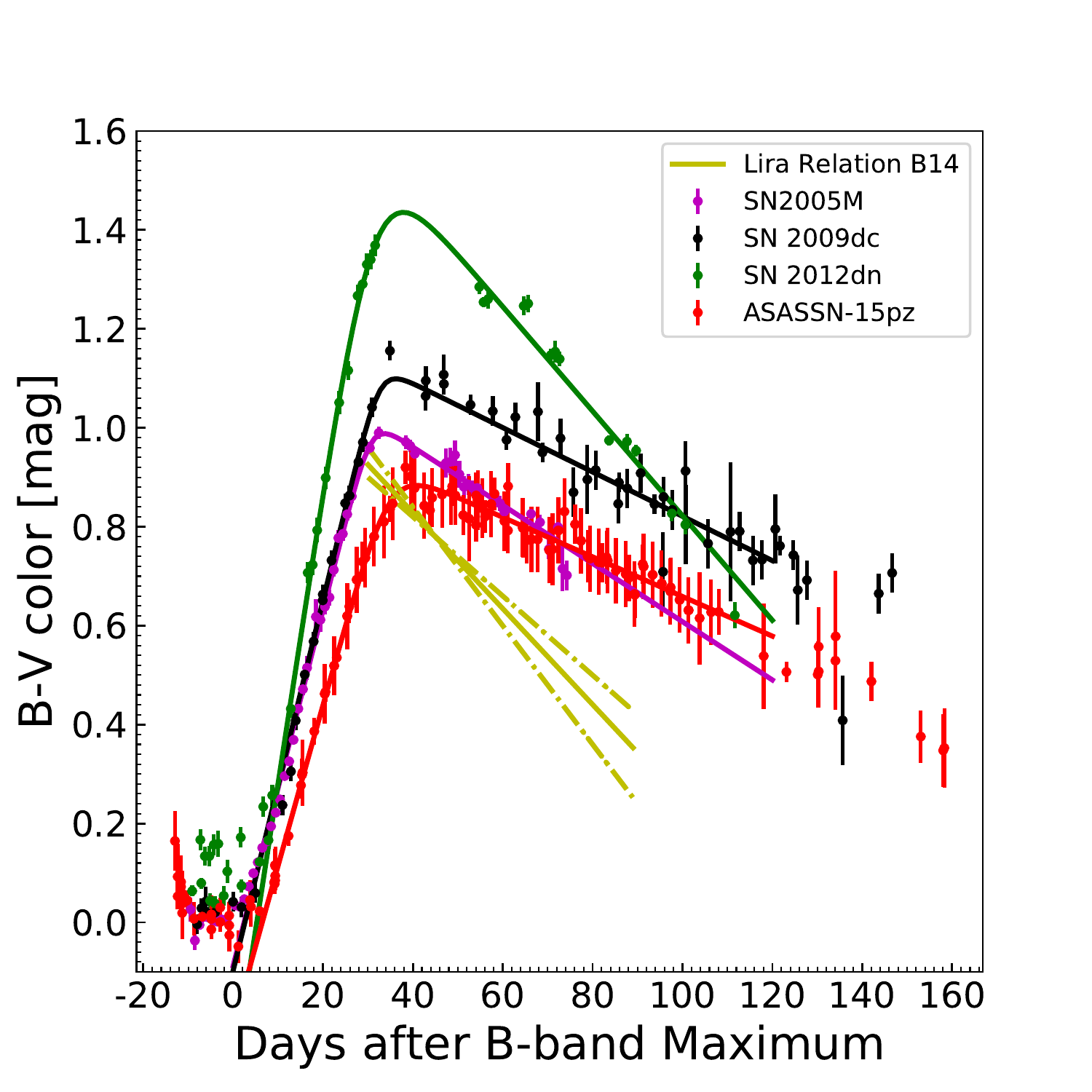}}
\caption{The ($B-V$) color of ASASSN-15pz (red) compared with other overly luminous SNe~Ia (SN~2009dc in black and SN~2012dn in green) and the 1991T-like SN~2005M in magenta. The mean and standard deviations of Lira relation from \cite{Burns2014} are shown as the yellow solid and dashed lines.}
\label{BV_lira}
\end{figure}

\citet{Burns2014} analyzed the intrinsic colors of SNe Ia using the CSP-I sample and introduced a new light-curve parameter -- ``color stretch" $s_{\rm BV} = (t^{\rm max}_{\rm B-V}/ 30$ days) where $t^{\rm max}_{\rm B-V}$ is the time of ($B-V$) color maximum after $B$-band maximum.  They found a tight correlation between $s_{BV}$ and peak luminosity (see the right panel of Fig.~\ref{fig:MB_m15_sbv}) for SNe Ia. This correlation between $s_{BV}$ and peak luminosity has significantly less scatter than that between $\Delta m_{15}(B)$ and peak luminosity for the low-luminosity SNe Ia ($M_B\gtrsim -19$). If we fit the ($B-V$) color evolution of ASASSN-15pz following the procedure of \citet{Burns2014}, we obtain  $t^{\rm max}_{\rm B-V} = 40.8\pm1.5$ days and $s_{\rm BV} = 1.36\pm0.05$. 

We also apply the same fitting procedure to SN~2009dc, SN~2012dn, and the 1991T-like SN Ia SN~2005M with the best-fit models shown in Fig.~\ref{BV_lira} and the parameters are given in Table~\ref{BVcolor_paras}.  For comparison, the Lira relation from \citet{Burns2014} and its standard deviation are also shown. The late-phase decline slope of ASASSN-15pz is similar to that of SN~2009dc, while slower than that of SN 2005M, which has a similar peak luminosity as ASASSN-15pz, and the Lira relation in general. We show in \S~\ref{as15pz_extinction} that the host-galaxy extinction of ASASSN-15pz is negligible, and the measured ($B-V$) color for ASASSN-15pz is thus intrinsic and can be used as an unreddened template for this sub-class.

\begin{table*}
\caption{Fitting parameters of $(B-V)$ color curve}
\begin{center}
\begin{tabular}{lccccc}
	\hline
	SN Name & $m_{15}(B)$ &$s_0$ &  $s_1$ & $t^{\rm max}_{B-V} (day)$ & $s_{\rm BV}$\\
	\hline
	2005M       & 0.87$\pm$0.02 & $0.036\pm0.001$ & $-0.0059\pm0.0003$ & $33.6\pm0.8$ & $1.12\pm0.03$\\
	2009dc      & 0.72$\pm$0.03& $0.037\pm0.003$ & $-0.0045\pm0.0005$ & $36.2\pm3.7$ & $1.21\pm 0.12$\\
	2012dn      & 0.92$\pm$0.04& $0.059\pm0.007$ & $-0.0106\pm0.0005$ & $37.9\pm2.5$ & $1.26 \pm0.08$\\
	ASASSN-15pz& 0.67$\pm$0.07&$0.033\pm0.003$ & $-0.0040\pm0.0002$ & $40.8\pm1.5$ & $1.36 \pm0.05$\\
	\hline
	\hline
\end{tabular}
\end{center}
\label{BVcolor_paras}
\end{table*}

\begin{figure}
\centerline{\includegraphics[width=10cm,height=10cm]{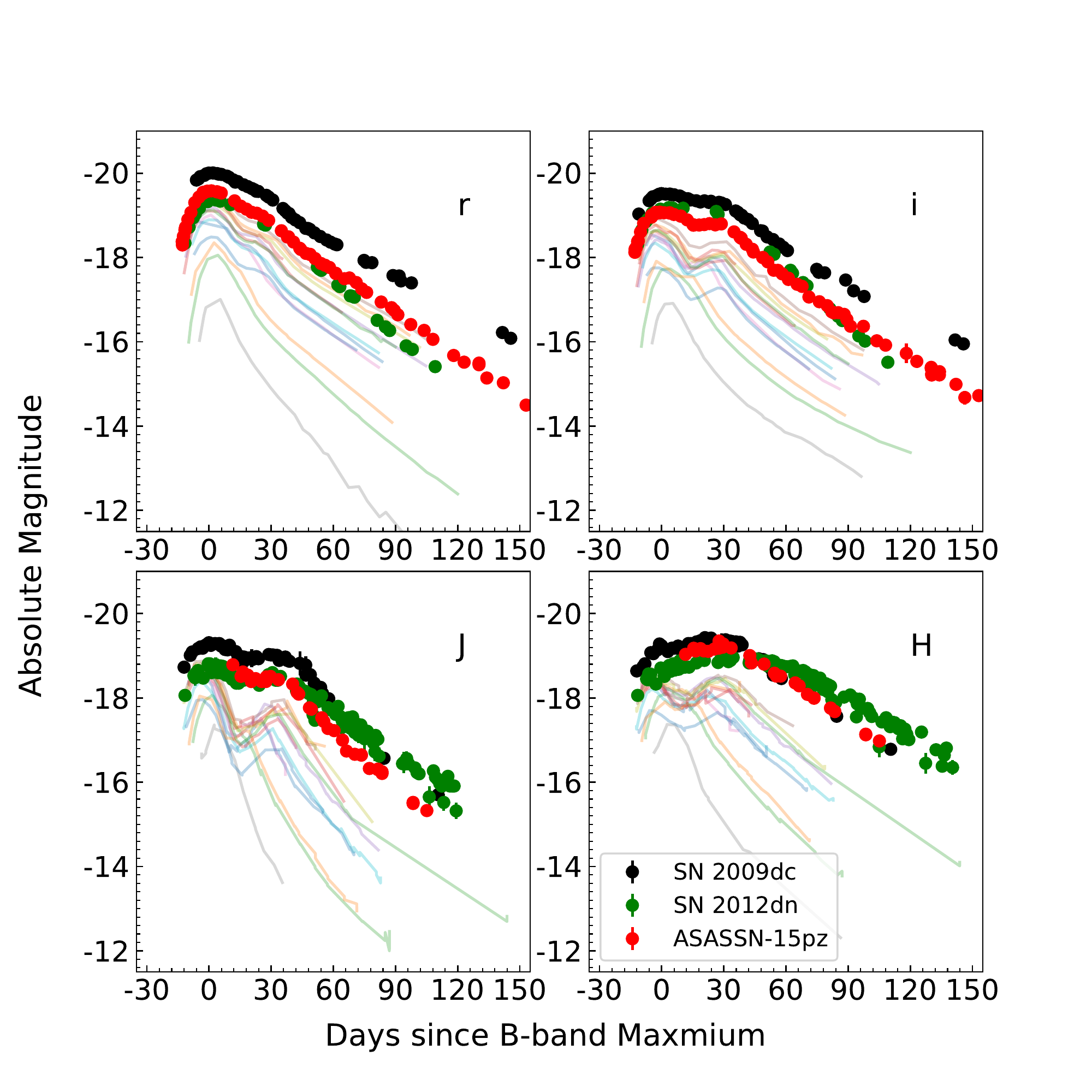}}
\caption{Absolute optical ($ri$) and NIR ($JH$)  light curves of
%bsolute magnitudes in SDSS $r$, $i$, $J$ and $H$ bands 
ASASSN-15pz compared to those of SN~2012dn \citep{Yamanaka2016}, SN~2009dc \citep{Taubenberger2011,Friedman2015},  and a  sample  selected from the literature \citep{Friedman2015, Krisciunas2017}.}
\label{NIR_lcs}
\end{figure}

In Fig.~\ref{NIR_lcs}, the $r$, $i$-,  $J$-, and $H$-band light curves of ASASSN-15pz are compared with those of SN~2009dc, SN~2012dn and the CSP-I SNe~Ia with good NIR light curves. Except for some low-luminosity SNe~Ia, the $iJH$-band light curves of SNe Ia (including the 1991T-like sub-class) generally exhibit prominent secondary maxima. ASASSN-15pz, SN~2009dc and SN~2012dn all have secondary maxima, but unlike the SNe~Ia 
from the CSP-I sample, they do not have significant troughs between the two peaks. Instead, their light curves are  ``shoulder-like''  between the two peaks with flat slopes. In the $H$ band, the peaks at $\gtrsim$30 days after the $B$-band maximum are broad and luminous, and they appear to overwhelm the peaks near the $B$-band peaks.

\subsection{\textit{Swift} UVOT light curve}
\label{UVOT_lcs_analysis}
\begin{figure}
\centerline{\includegraphics[width=10.5cm,height=9cm]{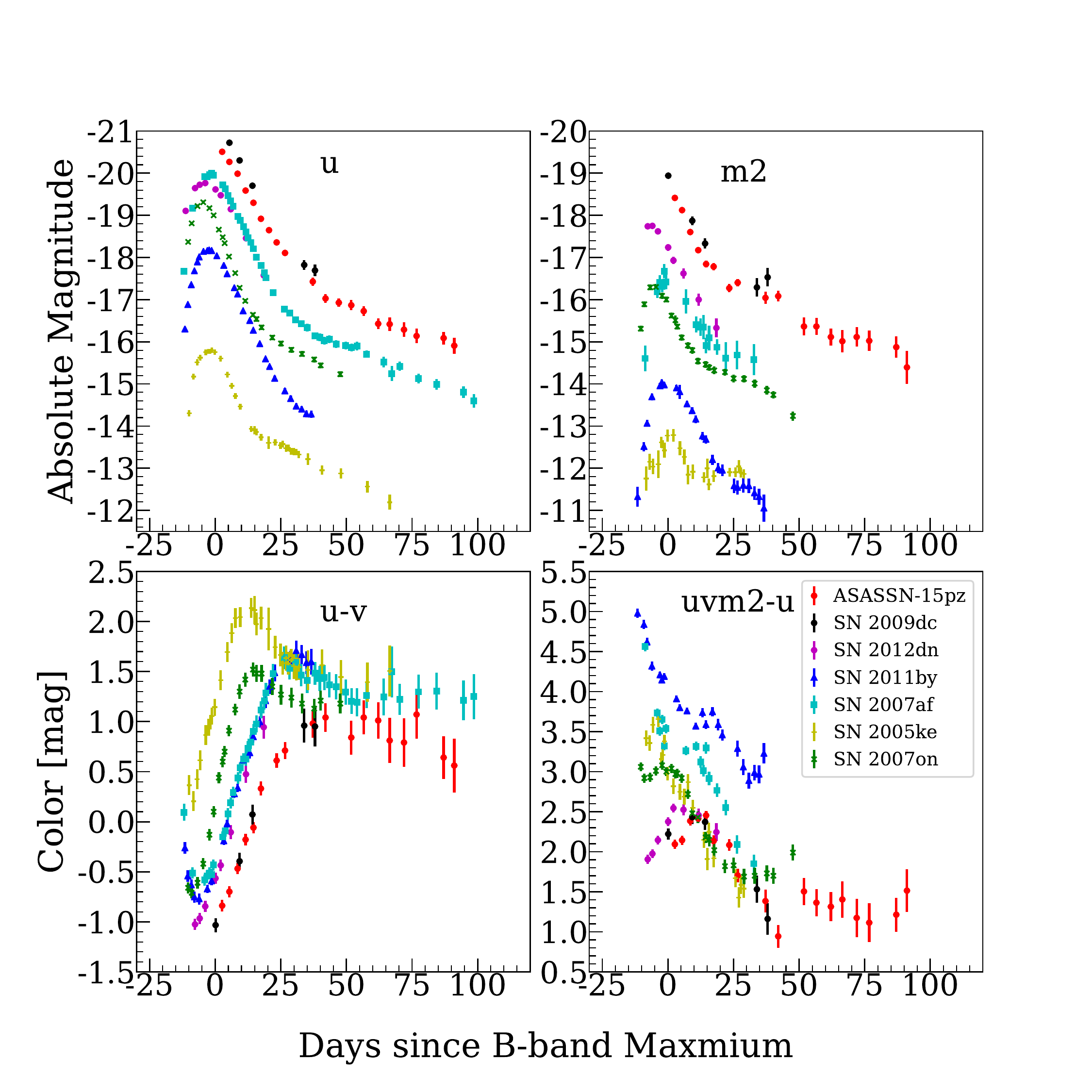}}
\caption{\textit{Swift} UVOT light curves of ASASSN-15pz compared to a sample of selected 2009dc-like SNe and SNe Ia with good UV coverage. Galactic and host galaxy extinction corrections are made for all targets. The top two panels show the absolute-magnitude light curves in the $u$ and $m2$ bands. The bottom two panels show the color evolution of ($u$-$v$) and ($uvm2$-$u$), which exhibit a significant diversity.}
\label{UV_lcs}
\end{figure}

In Fig.~\ref{UV_lcs}, we compare the evolution of ASASSN-15pz in $u$, $uvm2$, ($u-v$) and ($uvm2-u$) with SN~2009dc, SN~2012dn and a few SNe Ia with good UV coverage, including the normal SN~2007af \citep{Brown2009} and SN~2011by \citep{Milne2013}, the 1991bg-like SN~2005ke \citep{Brown2009}, and the ``transitional'' object SN 2007on \citep{Gall2018}, whose luminosity is between ``normal'' and 1991bg-like SNe~Ia. 
It is well known that 2009dc-like SNe are exceptionally UV bright around peak \citep[see, e.g.,][]{Brown14}, and ASASSN-15pz is also UV luminous. 
As shown in the bottom left panel of Fig.~\ref{UV_lcs}, the ($u-v$) colors of SNe Ia become redder after $B$-band maximum, then reverse and become bluer. Like the ($B-V$) color curves, the ($u-v$) color curves peak later for more luminous SNe. The ($uvm2-u$) color evolution of SNe Ia show more diversity, and ASASSN-15pz is the only object with $uvm2$ detections at late phases ($\gtrsim 50$\,d) to show a flat, late-time ($uvm2-u$) color evolution. ASASSN-15pz, SN~2009dc, and SN~2012dn are bluer in the UV than the SNe Ia in our comparison sample, suggesting a higher blackbody temperature or less Fe line blanketing for these objects. 

\subsection{Reddening and Luminosity}
\label{as15pz_extinction}
The Galactic extinction toward ASASSN-15pz is only $E(B-V)_{MW}=0.016$\,mag based on the \cite{Schlafly2011} recalibration of the \citet{Schlegel1998} dust map. The low Galactic extinction is consistent with the absence of Na I D absorption lines in all our spectra. \ion{Na}{1}~ D absorption is not detected at the host-galaxy redshift, with a 3\,$\sigma$ upper limit on the equivalent width of EW(\ion{Na}{1}~ D) $<{0.2}$\,\AA. This corresponds to a 1\,$\sigma$ upper limit of $E(B-V) < {0.04}$\,mag based on  \cite{Phillips2013}. 

Based on this nondetection of Na I D, we assume that the host-extinction correction for ASASSN-15pz is negligible. 
The reddening-free light curve of ASASSN-15pz makes it uniquely valuable to understand the intrinsic color and luminosity of luminous SNe-Ia-pec. 
As discussed in \S~\ref{optical_lcs} and shown in Fig.~\ref{BV_lira}, the post-peak ($B-V$) curve of SN~2009dc has a nearly identical slope to ASASSN-15pz ($s_1[{\rm 2009dc}] = -0.0045 \pm 0.0005$ and $s_1[{\rm ASASSN-15pz}] = -0.0040 \pm 0.0002$). In light of this and the close spectroscopic similarities between them, SN~2009dc and ASASSN-15pz likely have the same intrinsic color, which means that the host extinction for SN~2009dc is $E(B-V)_{\rm host} = 0.12\pm0.02\,{\rm mag}$ in addition to the Galactic extinction of $E(B-V)_{\rm MW}=0.062\,{\rm mag}$ \citep{Schlafly2011} for a total extinction of $E(B-V)_{\rm tot} = 0.18\,{\rm mag}$. 

{\citet{Silverman2011} obtained EW(\ion{Na}{1}~D) $=0.94\pm0.15$\,\AA\, for the host galaxy of SN~2009dc, which is consistent with the EW(\ion{Na}{1}~D) $\approx 1.0$\,\AA\, derived by \citet{Taubenberger2011}.} We note that according to the relation derived by \citet{Poznanski2012},  {EW(\ion{Na}{1}~D) $=0.94\pm0.15$\,\AA\, translates into $E(B-V)_{\rm host}=0.18^{+0.10}_{-0.07}\,{\rm mag}$}, which is consistent with our determination.

There is also non-negligible host-galaxy extinction for SN~2006gz and SN~2012dn, with EW(\ion{Na}{1}~D)[2012dn] $\approx0.75$ \,\AA\,\citep{Chakradhari2014} and EW(\ion{Na}{1}~ D)[2006gz] $\approx0.3$\,\AA\,\citep{Hicken2007}, respectively. For both SNe, the post-peak ($B-V$) decline rates are much steeper than ASASSN-15pz, thus we cannot derive their host reddening from a direct color curve comparison with ASASSN-15pz. \citet{Hicken2007} found that the post-peak (between 35 days and 51 days) ($B-V$) decline rate of SN 2006gz was similar to SNe Ia, and by applying the Lira relation, they derived $E(B-V)=0.18\pm{0.05}{\rm\, mag}$, which was consistent with the limit of $E(B-V)_{\rm host} \leq0.15\pm{0.08}{\rm\, mag}$ that they estimated using EW(\ion{Na}{1}~ D). However, there is no evidence supporting the idea that the intrinsic color evolution of luminous SNe~Ia-pec follow that of normal SNe~Ia. In fact,  \citet{Chakradhari2014} found that the Lira relations would yield $E(B-V)_{\rm Lira}\approx0.43{\rm\, mag}$ for SN~2012dn, which would be much higher than the $E(B-V)_{\rm host}\approx0.12{\rm\, mag}$ estimated from the \ion{Na}{1}~D absorption. Because we have no extinction-free analogs of SN 2006gz or SN 2012dn, it is not safe to derive their host-reddening by making assumptions on their intrinsic color evolutions. We adopt extinction estimates derived from the \ion{Na}{1}~D absorption for SN~2006gz and SN~2012dn in our analysis. 

The adopted extinction and peak luminosities for these four luminous SNe~Ia-pec are given in Table~\ref{09dc_paras}, which also includes the parameters for SN~1991T \citep{Phillips1999}. The peak luminosity of ASASSN-15pz is very similar to SN~1991T but $\approx 0.6$\,mag dimmer than SN 2009dc.

\begin{table*}
\caption{Basic comparison between 2009dc-like SNe~Ia and SN~1991T}
\begin{center}
\begin{tabular}{ccccccc}
	\hline
SN                    &   z       & $\mu$(mag)  & $E(B-V)_{\rm MW}$(mag) & $E(B-V)_{\rm host}$(mag)  & $M_B^{\rm peak}$(mag) & $M_V^{\rm peak}$(mag) \\
	\hline
ASASSN-15pz &0.014837   & 33.85 &   0.016     & 0.0               &  $-19.69$    & $-19.67$ \\
2012dn       &0.010187  &33.14   &   0.06       & 0.12            &   $-19.51$    & $-19.40$ \\
 2009dc       & 0.021391  & 34.85 &   0.06       & 0.12            &   $-20.26$    & $-20.22$ \\
2006gz       & 0.0234     &35.07   &   0.07       & 0.18            &   $-20.03$    & $-19.86$ \\
1991T        &0.00579     &30.74   &   0.019    & 0.15            &   $-19.73$    &  $-19.74$ \\
2005M       & 0.024624  & 35.09   & 0.027      & 0.06            &    $-19.45$  & $  -19.36 $\\
	\hline
	\hline
\end{tabular}
\end{center}
\label{09dc_paras}
\end{table*}

\subsection{Bolometric light curve}
\label{blc}
\begin{figure}
\centerline{\includegraphics[width=10cm,height=6cm]{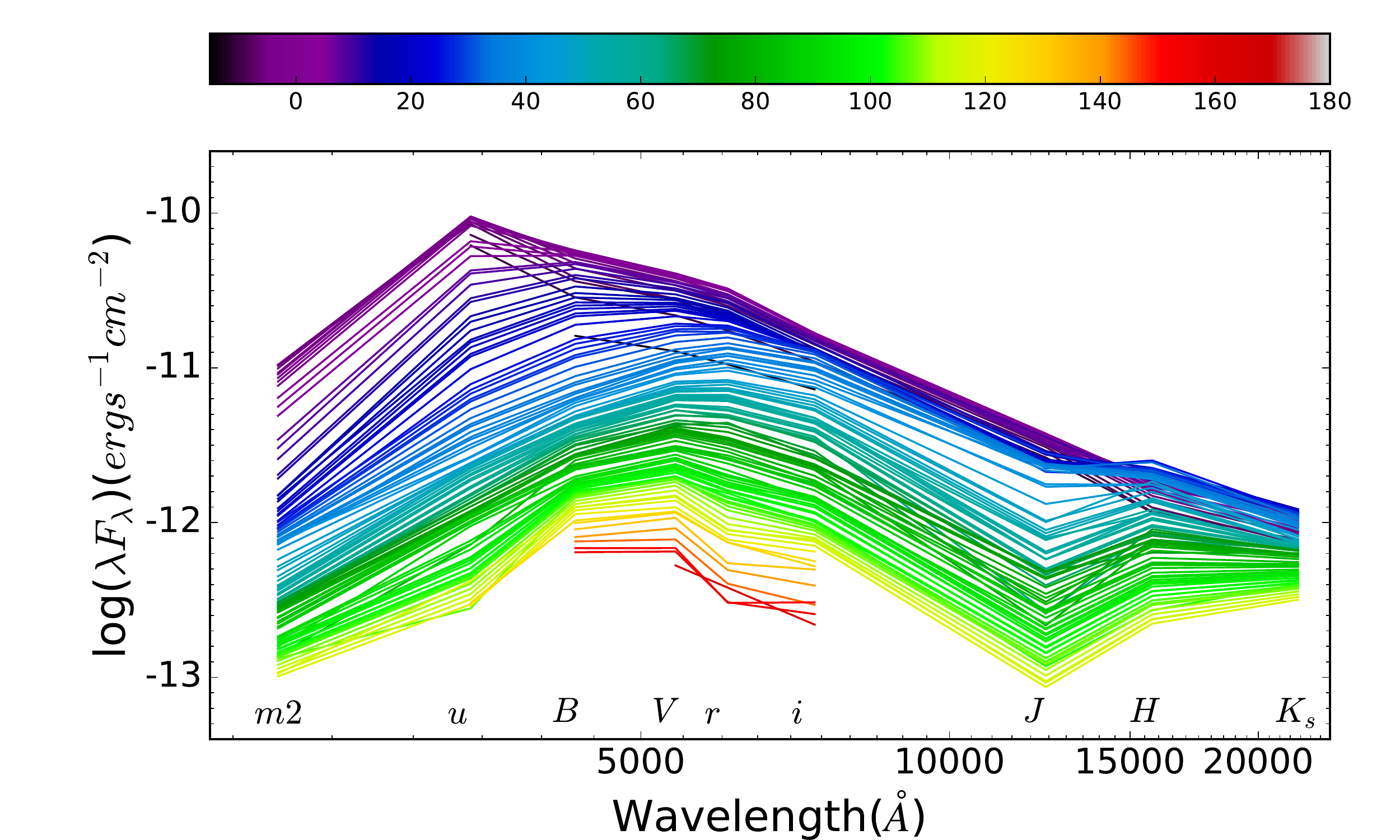}}
\caption{Rest-frame SEDs of ASASSN-15pz. The colors indicate the phase relative to the $B$-band maximum as indicated by the color bar.}
\label{sed}
\end{figure}

\begin{figure}
\centerline{\includegraphics[width=10cm,height=12.5cm]{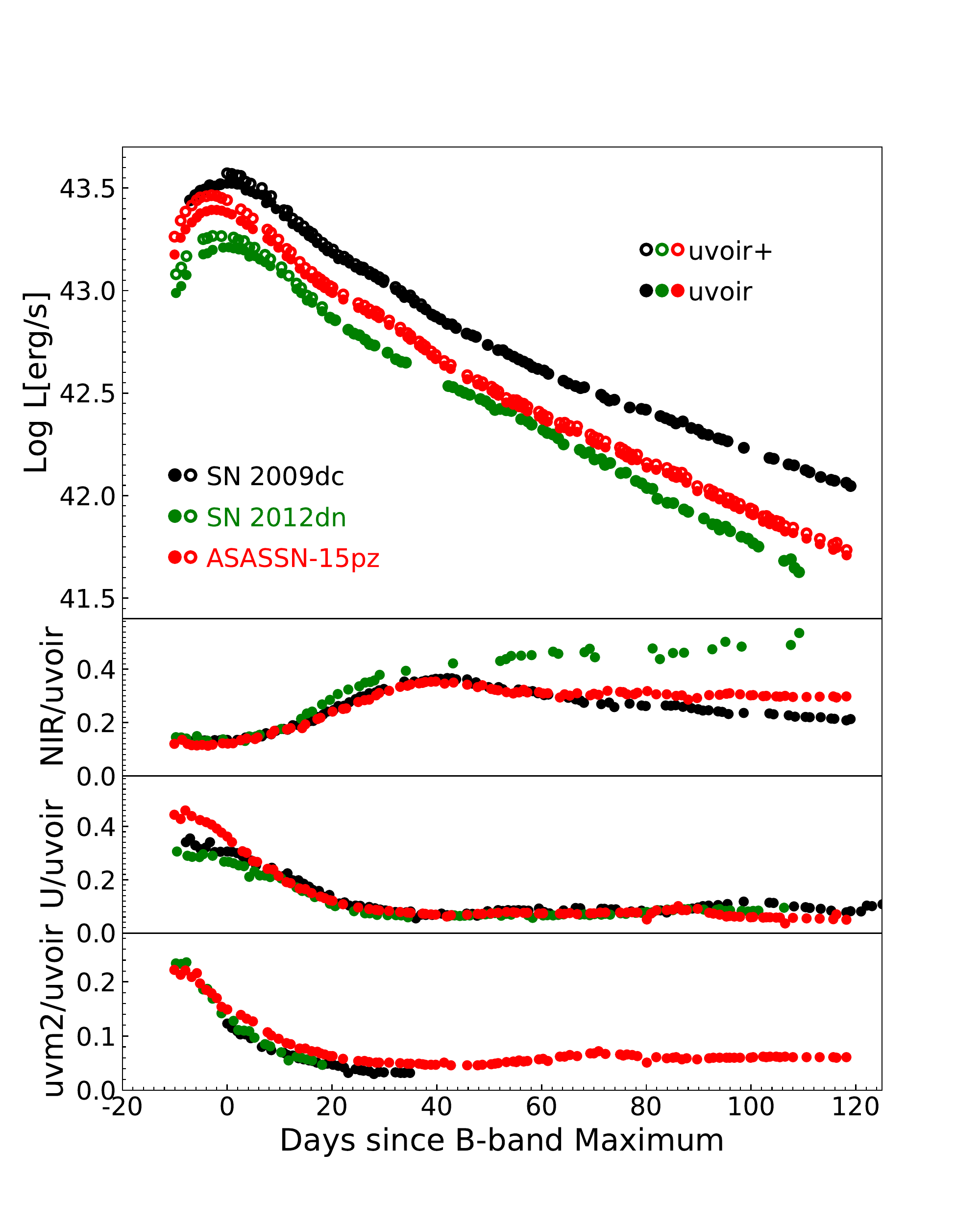}}
\caption{Bolometric light curve and fractional contributions from different wavelength ranges to ASASSN-15pz, SN~2009dc, and SN~2012dn. In the uppermost panel, the uvoir bolometric light curves shown in filled circles (ASASSN-15pz in red, SN~2009dc in black, and SN~2012dn in green) are constructed from UVOT $u$ (or ground-based $U$ band), $BVri$($RI$) and $JHKs$ bands photometry data, while the uvoir$+$ bolometric lights using empty circles include UVOT $uvm2$-band flux. The bottom three panels show the fractional contributions from the NIR (from 8150 to 23550~\AA), $U$ (from 3100 to 3930~\AA), and $uvm2$  (from 2000 to 3000~\AA) bands relative to the uvoir luminosity. }
\label{bol_lcs}
\end{figure}

To construct a bolometric light curve we first convert the observed UV, optical, and NIR magnitudes into monochromatic fluxes and de-redshift them into the rest frame. We use only the $uvm2$- and $u$-band {\textit Swift} data because of the red-leak issue in the $uvw2$ and $uvw1$ filters \citep{Brown2010}. The resulting spectral energy distributions (SEDs) are shown in Fig.~\ref{sed}, where the peak wavelength changes gradually from blue to red as the ejecta expand and cool. Since the UV and NIR observations started about 2 days and 10 days after peak, respectively, we extrapolate our early-phase UV and NIR light curve to construct the bolometric light curves using the shapes of the UV and NIR light curves of SN~2012dn and SN 2009dc, respectively. We integrate the SEDs to construct pseudo-bolometric (``uvoir$+$'') light curves covering from 2000 to 23550~\AA. We also construct pseudo-bolometric (``uvoir'') light curves covering 3100--23550 \,\AA, in order to make direct comparisons with those for SN~2009dc and SN~2012dn (see Fig.~\ref{bol_lcs}). These pseudo-bolometric light curves of SN~2009dc and SN~2012dn are constructed using the same method as for ASASSN-15pz. 

For SNe~Ia, the flux at wavelengths blueward of 3100~\AA, is generally believed to make a small contribution to the bolometric flux \citep[see, e.g.,][]{Contardo2000, Contreras2018}. But SN~2009dc and SN~2012dn are exceptionally blue \citep{Brown2014} and the UV contribution to their bolometric luminosity near the peak is significant. Similarly, as shown in the bottom two panels of Fig.~\ref{bol_lcs}, we find that for ASASSN-15pz, there is significant UV contribution to the bolometric light, and the U/uvoir (MUV/uvoir) fraction evolves from as high as  $ \sim 45\%$ ($ \sim 23\%$) at $-$12 days to $\sim 38\%$($\sim 15\%$) at the $B$-band peak, and then to a constant level of $\sim 10 \%$ ($ \sim 6\%$) for +30 days and the later phases. The early-time UV and late-time NIR contributions show considerable diversities reflecting the diversities seen in the color evolution discussed in Sections \ref{UVOT_lcs_analysis} and \ref{optical_lcs}. This makes it important to have UV and NIR data to construct a reliable bolometric light curve rather than assuming the fractions observed from some template objects.

\subsection{Mass of $^{56}$Ni}

SNe Ia are powered by the radioactive decay of $^{56}$Ni synthesized in the thermonuclear explosion, and the mass of $^{56}$Ni is a key physical parameter. We can approximately estimate the mass of $^{56}$Ni using ``Arnett's rule'' \citep{Arnett1982} that $L_{\rm bol}^{\rm peak} \approx \alpha Q{(t_{\rm peak})} \: M_{Ni}$, 
where $Q(t_{\rm peak})$ is the rate of radioactive energy production per unit nickel mass at time of peak luminosity after the explosion, $\alpha$ is a coefficient of order unity and $L_{\rm bol}^{\rm peak}$ is the peak bolometric luminosity. The ASAS-SN detection on UT 2015 September 27.16 means that $t_{\rm peak}>14.6\,{\rm d}$. If we fit our early light curve (phase $<-10{\rm d}$) to a fiducial fireball model $f=A(t+t_R)^2$ we get a best-fit estimate of $t_{\rm peak}=21.4\,{\rm d}$. For comparison, \cite{Silverman2011} obtained a direct constraint on $t_{\rm peak}$ for SN 2009dc from early detections as $t_{\rm peak}>21$ days, and combining with non-detections they estimated a rise time of $23$ with an uncertainty of 2 days.  The  $^{56}$Ni mass estimate using  ``Arnett's rule'' is sensitive to the estimate of explosion time, and we adopt an uncertainty of 2 days similar to \cite{Silverman2011}. With $L_{\rm bol, peak}= 2.7 \times 10^{43} $ erg s$^{-1}$ for ASASSN-15pz, we obtain $M_{^{56}{\rm Ni}} \approx 1.1 \pm 0.4$ M$_{\astrosun}$ with $t_{\rm peak}=21.4\pm2$ days and $\alpha=1.0 \pm 0.2$ \citep[see, e.g.,][for various estimates of $\alpha$]{Branch1992, Stritzinger2005}. 

\begin{figure}
\centerline{\includegraphics[width=10cm,height=12.5cm]{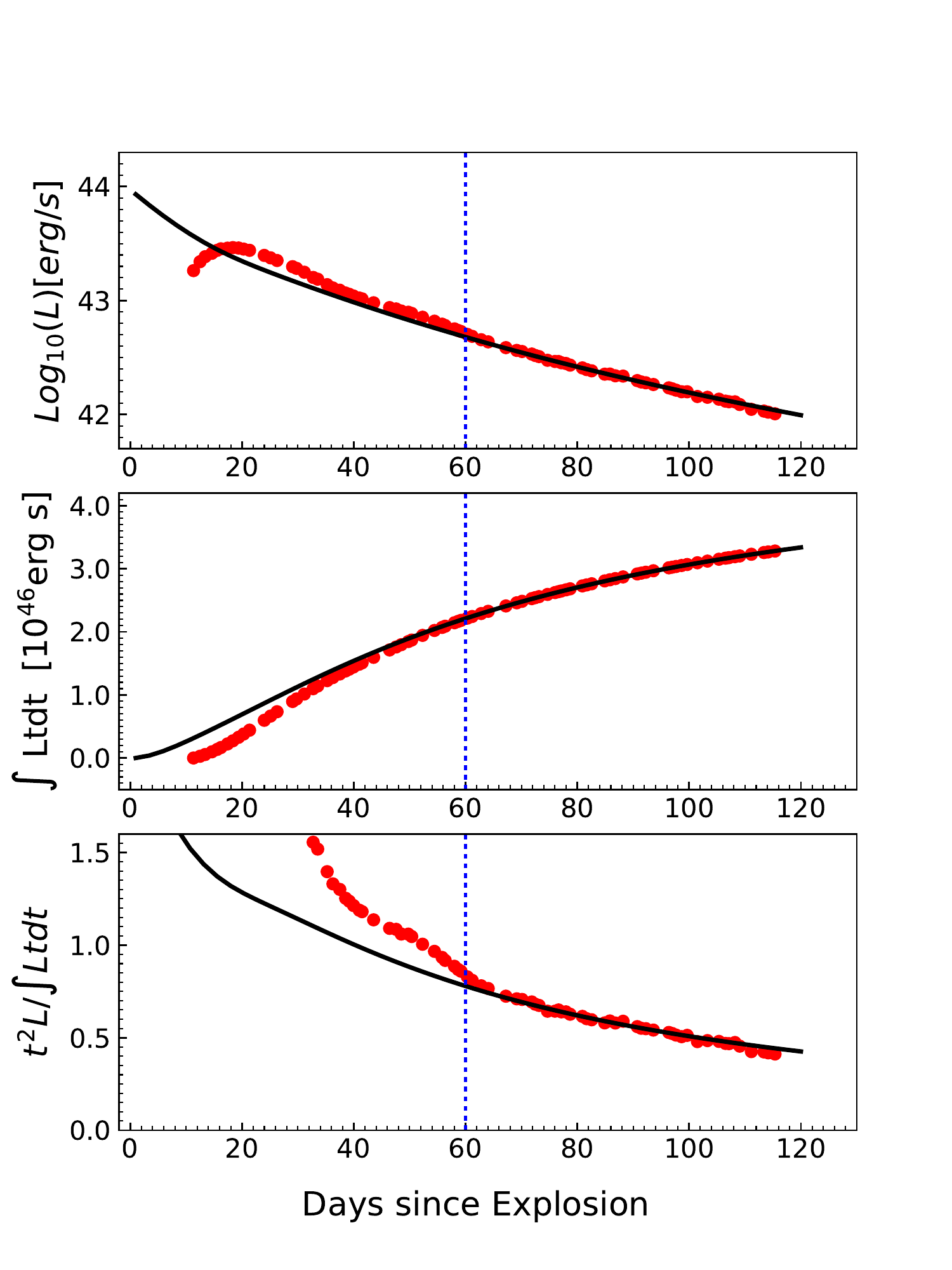}}
\caption{Model fit (black solid line) to the observed bolometric light curve of ASASSN-15pz. The bottom panel shows the fitting result to ${t^2L}/{\int Ltdt}$ and the middle panel shows the fitting result to the integral $\int Ltdt$. The black line in the upper panel represents the energy deposition of the gamma rays and positrons produced by the decay of $^{56}$Ni for the best-fit $^{56}$Ni mass and  $\gamma$-ray escape timescale $t_0$. The dotted lines indicate the phase (60 days) after which the data are used for the fits.}
\label{bol_fitting}
\end{figure}

The radioactive-decay energy is principally released as $\gamma$-rays and positrons. 
At late times, the fraction ofÂ $\gamma$-ray energy deposited in the ejecta is given by $t^2_0/t^2$, where $t_0$ Â is the Â $\gamma$-ray escape timescale \citep[e.g.,][]{Jeffery1999, Stritzinger2006}. For $t\ll t_0$, all gamma-rays are trapped and the deposition fraction is unity. To a good approximation, the deposition can be interpolated by $1-\exp(-t_0^2/t^2)$ at all times \citep[see, e.g.][]{Wygoda2019a}.
As shown by \citet{Katz2013}, at sufficient late times, there are two relations between the deposition and the bolometric luminosity, $Q(t)=L(t)$ and the integral $\int_0^tQ(t')t'dt' = \int_0^tL_{bol}(t')t'dt'$, and the $^{56}$Ni mass can be derived using these relations \citep[see some examples by][]{Blondin2018, Wygoda2019a}.
For ASASSN-15pz, we first obtain $t_0 = 46.4\pm0.5$ days by fitting ${t^2L}/{\int Ltdt}$, which is independent of $^{56}$Ni mass, to the observed bolometric light from 60 days to 115 days after explosion, as shown in bottom panel of Fig.~\ref{bol_fitting}. Then we fit the integral $\int Ltdt$, as shown in the middle panel, and obtain $M_{^{56}Ni} = 1.13 \pm 0.14$ M$_{\astrosun}$. Unlike Arnett's rule, the estimate of $M_{^{56}Ni}$ with the \citet{Katz2013} integral method is insensitive to the exact value of $t_{\rm peak}$, as the integral $\int Ltdt$ has little weight from the early phases. The $\gamma$-ray escape timescale $t_0$ depends on the $^{56}$Ni mass weighted column density of the ejecta \citep[see, e.g.,][]{Jeffery1999,Wygoda2019a}, and {$t_0 = 46.4\pm0.5$d}  for ASASSN-15pz is higher than almost all SNe Ia in the sample analyzed by \citet{Wygoda2019a}, suggesting that it has a relatively high ejecta column density. The same method applied to our bolometric light curve for SN 2009dc yields $t_0=55.1\pm1.0$ days and $M_{^{56}Ni} = {1.6 \pm 0.1 M_{\astrosun}}$, implying higher ejecta column density and $M_{^{56}Ni}$ than ASASSN-15pz.

The broad consistency between the two $^{56}$Ni mass estimates is similar to that found for SNe~Ia and supports radioactive decay as the energy source in these objects over alternative scenarios such as ejecta-CSM interaction \citep[see, e.g.,][]{Noebauer2016}.

%\newpage
\section{Optical Spectra}
\label{spectra}

\begin{figure*}
\centerline{\includegraphics[width=16cm,height=16cm]{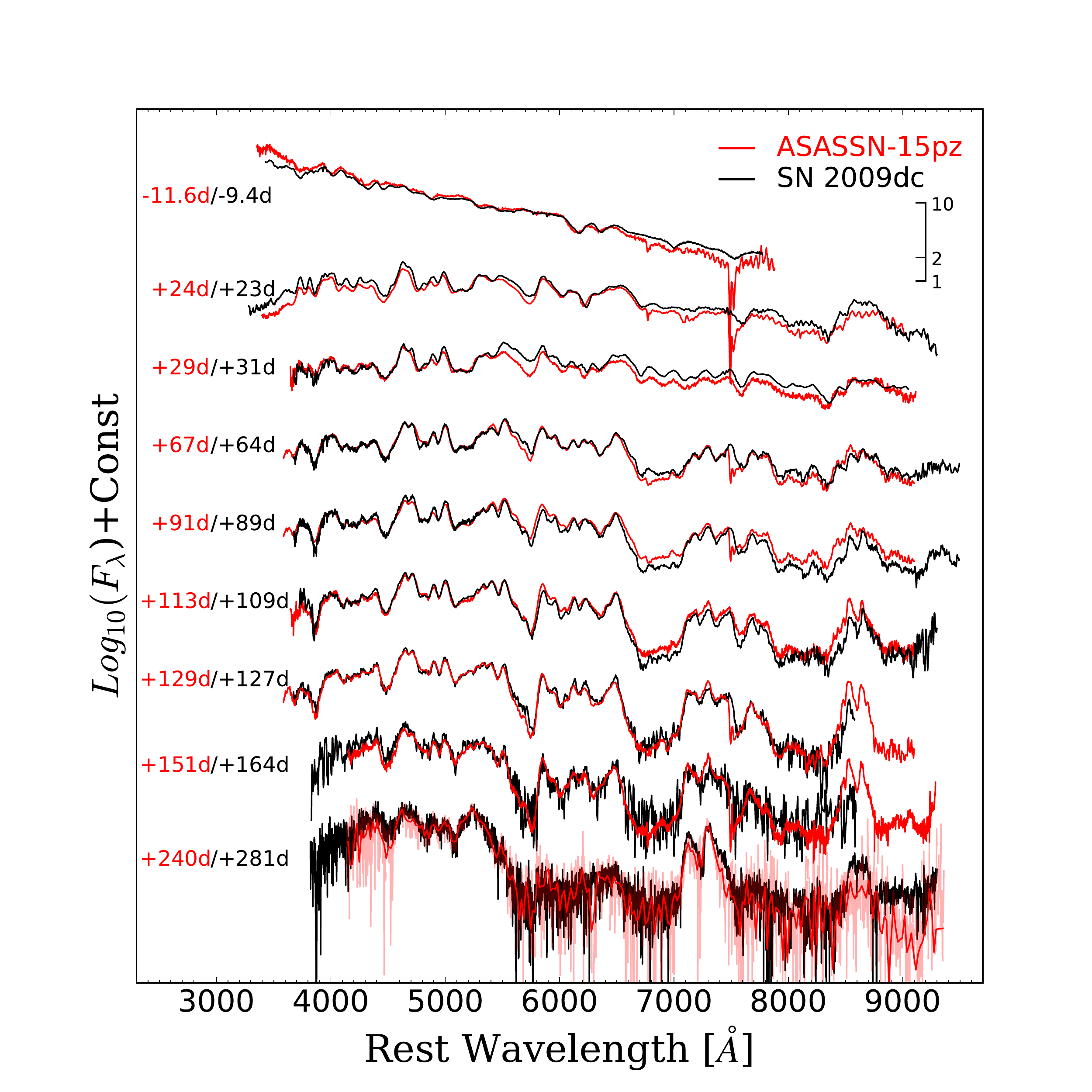}}
\caption{Spectra of ASASSN-15pz (red) compared to those of SN~2009dc (black), revealing that they are  spectroscopic twins  from pre-peak to nebular phases (phases are labeled at the left). The Y-axis is on logarithmic scale, and the vertical bar on the upper right shows the scale for factors of 2 and 10. All spectra have been shifted to the rest frame, and the median fluxes of SN~2009dc's spectra between 5000 and 5500\,\AA\ have been normalized to match those of ASASSN-15pz at the corresponding epochs.  All reduced spectra are made available at  the Weizmann Interactive Supernova Data Repository (WISeREP; \citealt{2012PASP..124..668Y}).}
\label{as15pz_twin}
\end{figure*}

In Fig.~\ref{as15pz_twin}, we compare rest-frame spectra for ASASSN-15pz ranging from $-11.6$ days to $+240$ days  with those of SN~2009dc at similar phases. The spectra are extinction-corrected. At all phases, the spectral features of ASASSN-15pz are almost identical to SN 2009dc: ASASSN-15pz is a spectroscopic twin of SN~2009dc. In the following sections, we discuss the spectroscopic features at various phases.

\subsection{Pre-maximum spectrum}

\begin{figure}
\centerline{\includegraphics[width=10cm,height=10cm]{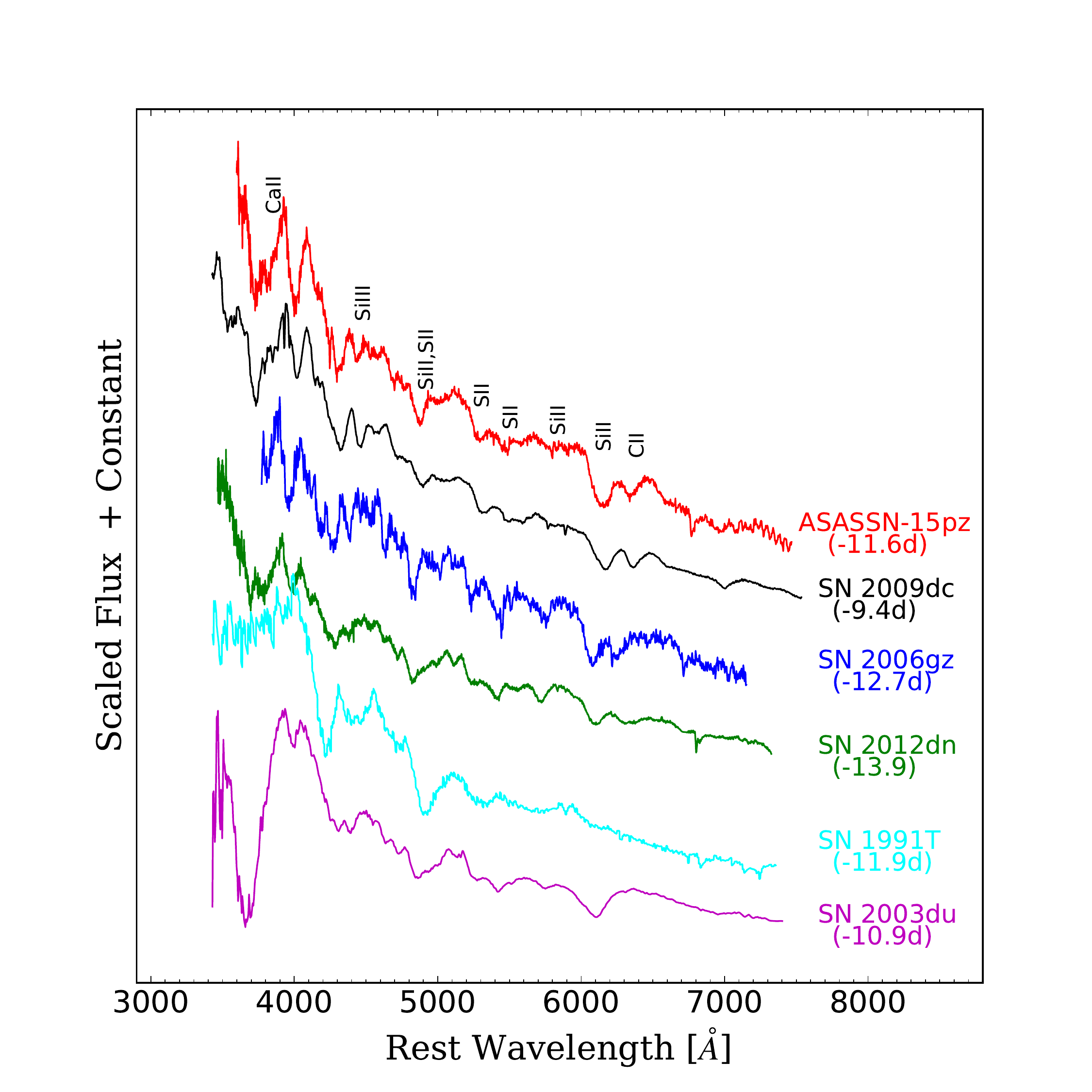}}
\caption{Pre-maximum spectrum of ASASSN-15pz compared with various SNe~Ia at similar phase (in parentheses). From top to bottom, the SNe are over-luminous SN~2009dc \citep{Taubenberger2011}, SN~2006gz \citep{Hicken2007}, SN~2012dn  \citep{Parrent2016}, the over-luminous prototype SN~1991T  \citep{Mazzali1995}, and the normal SN~2003du.}
\label{pre-maximum}
\end{figure}

\begin{figure}
\centerline{\includegraphics[width=9cm,height=9cm]{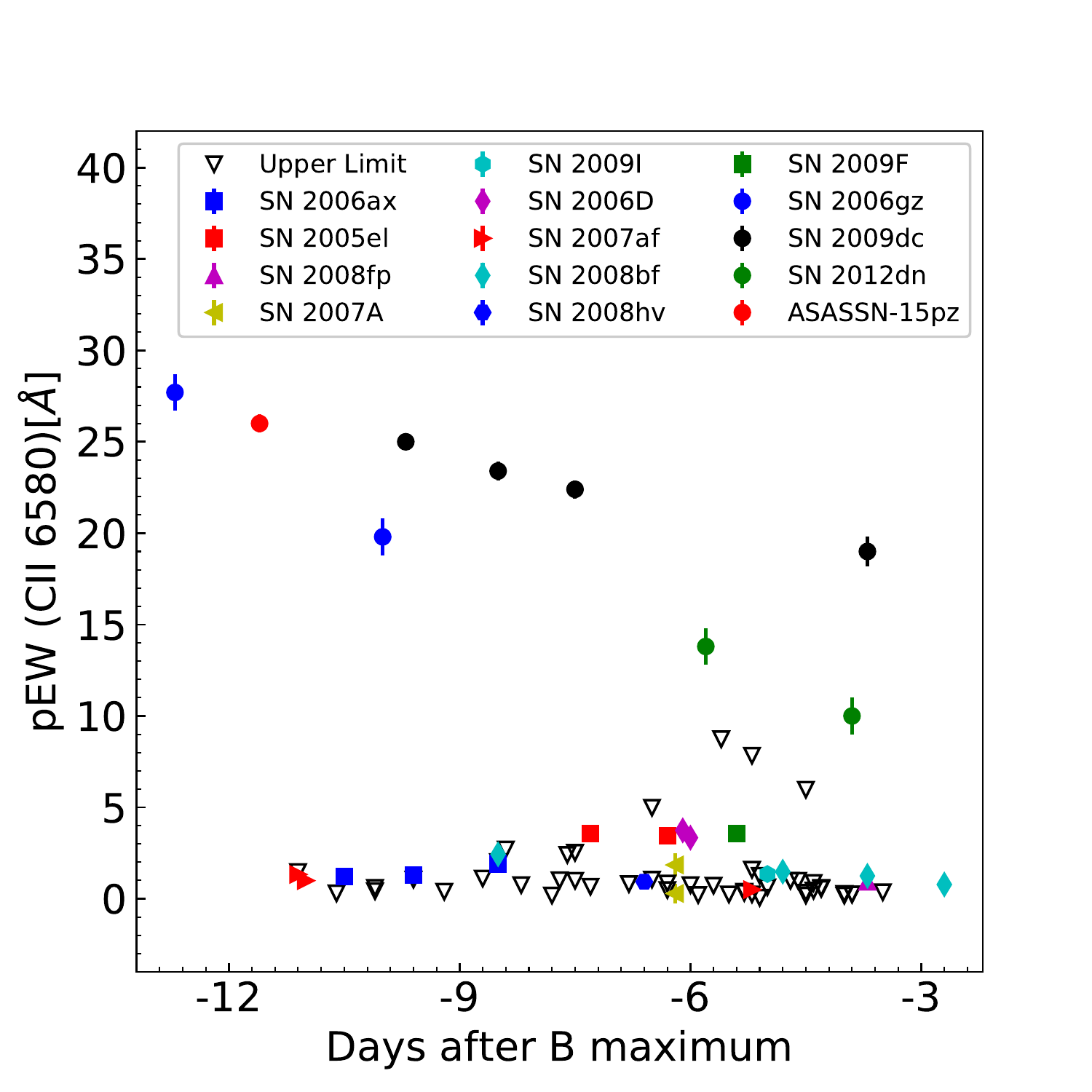}}
\caption{Pseudo-equivalent width of \ion{C}{2}  $\lambda 6580$ in ASASSN-15pz as compared to SN~2009dc, SN~2006gz, SN~2012dn and other SNe~Ia from the CSP-I sample \citep{Folatelli2012}.}
\label{CII6580_pEW}
\end{figure}

\begin{figure}
\centerline{\includegraphics[width=9cm,height=9cm]{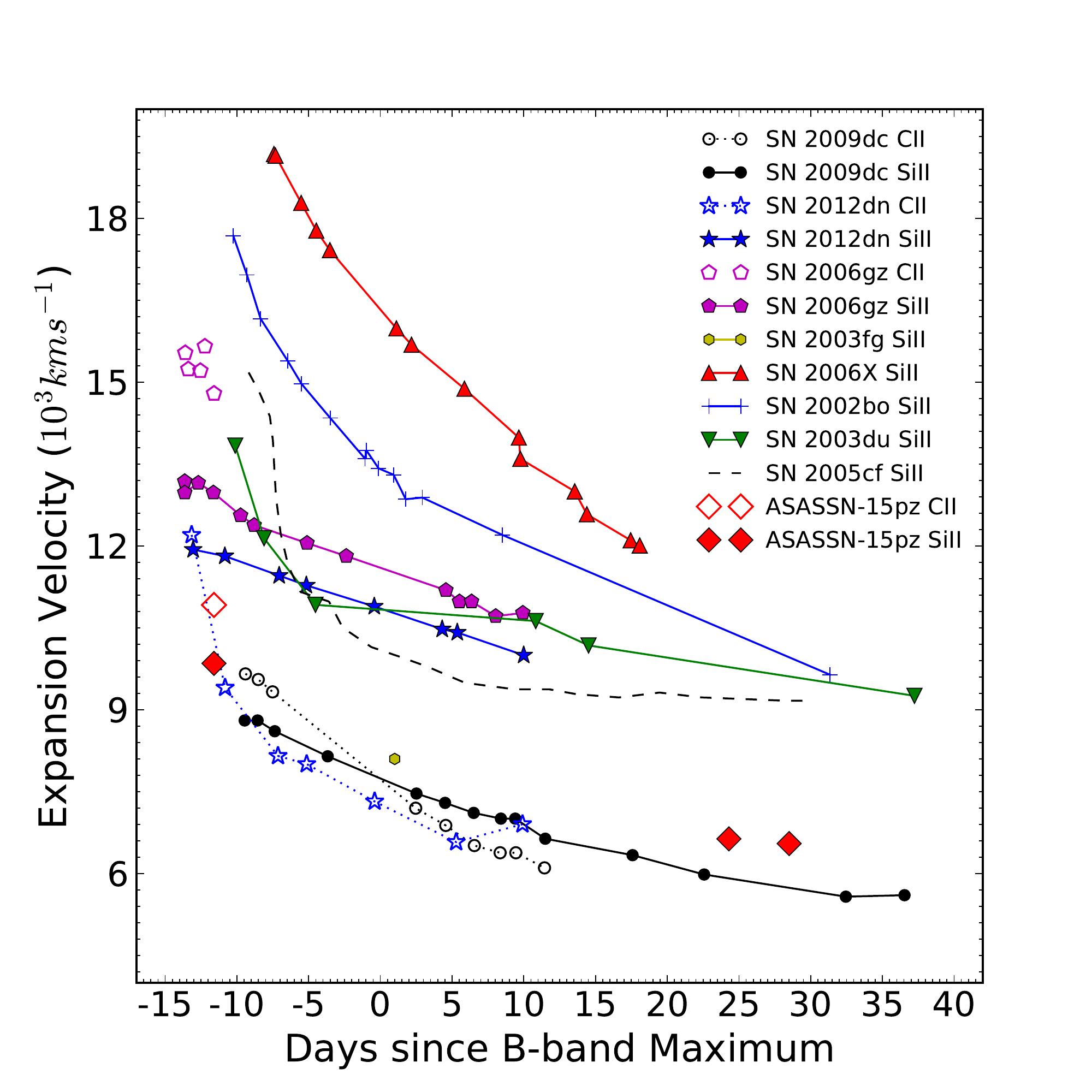}}
\caption{Spectral velocity evolution of the \ion{Si}{2} $\lambda 6355$ absorption line for ASASSN-15pz as compared with those of SNe Ia-pec (SN 2009dc, SN 2012dn, SN 2006gz, SN 2007if, 2003fg), and SNe Ia (SN 2006X, SN 2002bo, SN 2003du and SN 2005cf). \ion{C}{2} $\lambda 6580$ velocities are shown for ASASSN-15pz, SN 2009dc, SN 2006gz, and SN 2012dn.}
\label{velocity}
\end{figure} 

\begin{figure*}
\centerline{\includegraphics[width=18cm]{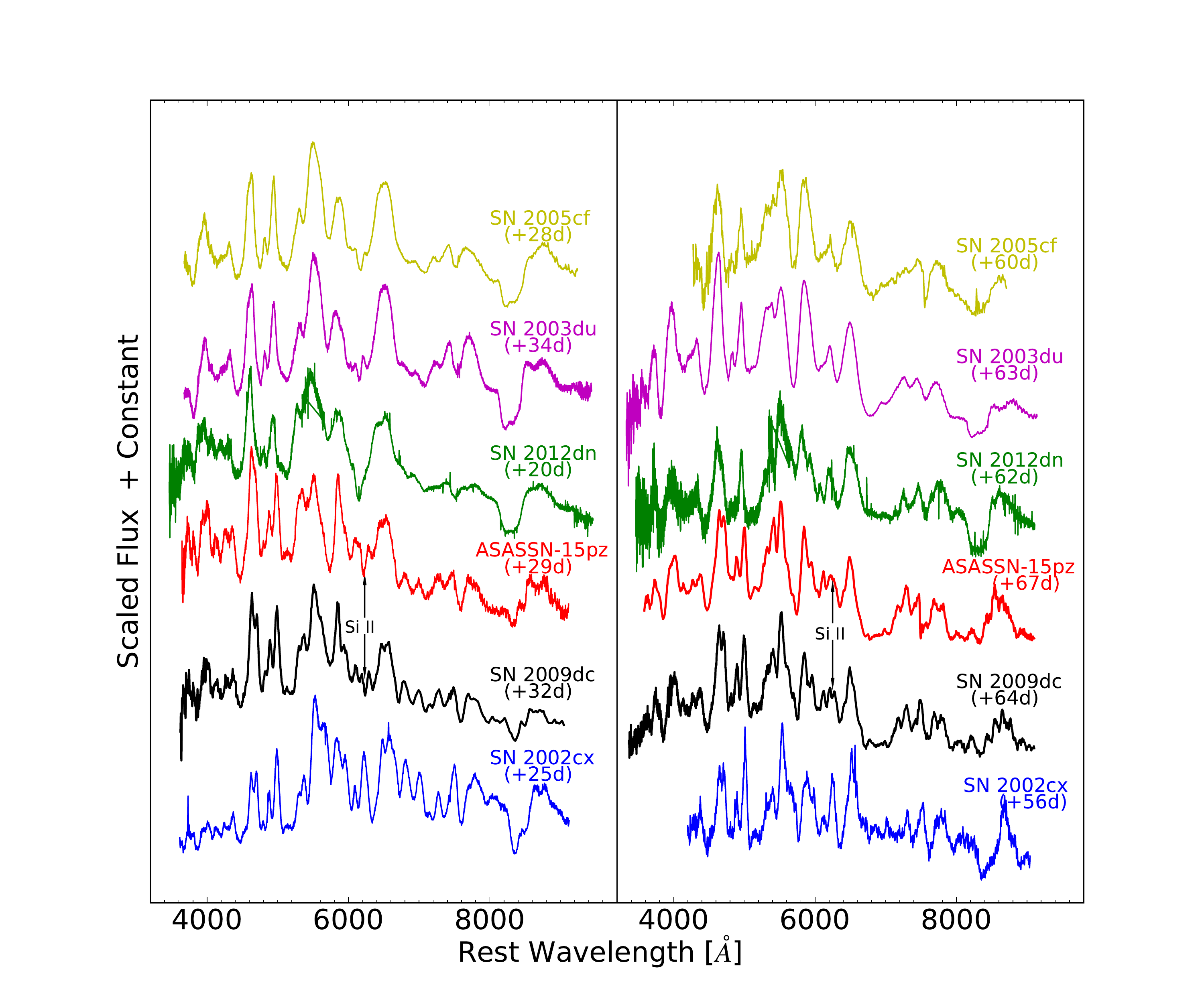}}
\caption{Spectra of ASASSN-15pz at phase +28.5 days (in left panel) and phase +67 days (in right panel) are compared with those of SN 2012dn, SN 2009dc, SN 2002cx, SN 2005cf, and SN 2003du. The phases relative to maximum light are indicated for each spectrum (in parentheses).}
\label{one_two_month_spectra}
\end{figure*}

Using the earliest spectrum shown in Fig.~\ref{as15pz_twin}, \cite{atel_classification} reported their spectroscopic classification of ASASSN-15pz as a SN~Ia at a phase near maximum. There is a prominent \ion{C}{2} absorption feature redward of the \ion{Si}{2} $\lambda$6355 feature, distinguishing it from pre-peak spectra of the Type~Ia SNe 2003du \citep{Stanishev2007} and 1991T  \citep{Filippenko1992_91T,Jeffery1992,Ruiz1992,  Mazzali1995}, as shown in Fig.~\ref{pre-maximum}. This strong \ion{C}{2} feature appears to be a unique signature of 2009dc-like objects at early phases (e.g., SN~2009dc from \citealt{Taubenberger2011}; SN~2006gz from \citealt{Hicken2007}; SN~2012dn from \citealt{Parrent2016}).

The  \ion{C}{2} absorption feature is commonly interpreted as being due to unburned carbon, and the \ion{C}{2} species can be seen at wavelengths of 4267, 4745, 6580 and 7234\,\AA\,\  in the optical spectra \citep[see, e.g.,][]{Mazzali2001}.  \citet{Folatelli2012} found that at least 30$\%$ of CSP-I SNe~Ia exhibit \ion{C}{2} absorption at $\lambda$6580, and a similar fraction was reported by \citet{Parrent2011}. The \ion{C}{2} feature is unusually strong in 2009dc-like objects. This is shown in Fig.~\ref{CII6580_pEW}, where the pseudo-equivalent width (pEW) of \ion{C}{2}  $\lambda$6580 of ASASSN-15pz is shown along with those of  SN 2006gz, SN~2009dc, and SN~2012dn in comparison to SNe~Ia from the CSP sample in \cite{Folatelli2012}. The absorption strengths of 2009dc-like objects appear to separate them from the CSP SNe~Ia.

We measure expansion velocities of  9850\,{\rm km\,s}$^{-1}$ and 10900\,{\rm km\,s}$^{-1}$ from the \ion{Si}{2} $\lambda 6355$ and \ion{C}{2} $\lambda 6580$ absorption lines in the $-11.6$ day spectrum. The expansion velocities of ASASSN-15pz are generally $\sim 1000\,{\rm km\,s}^{-1}$ higher than SN~2009dc at similar phases but they are substantially lower than normal SNe Ia, as illustrated by SN 2002bo \citep{Benetti2004}, SN 2003du \citep{Stanishev2007}, SN 2005cf \citep{Wang2009bf}, and SN 2006X \citep{Yamanaka2009b} in Fig.~\ref{velocity}. The slowly expanding ejecta are also a well known feature of the 2009dc-like SNe.

\subsection{One and two months after Maximum}

In Fig.~\ref{one_two_month_spectra}, we compare the spectra of ASASSN-15pz one and two months after peak with those of the peculiar SN Ia 2002cx \citep{Li2003}, SNe Ia SN 2003du \citep{Stanishev2007} and SN 2005cf \citep{Wang2009bf},  SNe Ia-pec SN 2009dc  \citep{Taubenberger2011}, and SN 2012dn \citep{Childress2016,Yamanaka2016} at similar phases. All pronounced line features in SNe Ia (here SN 2003du and SN 2005cf) are present in ASASSN-15pz, but the features for ASASSN-15pz (like 2009dc) are narrower and sharper. 
 \ion{Si}{2} $\lambda 6355$ absorption was detected in the +64 days spectrum of SN 2009dc, and it is tentatively detected in the +67 days spectrum of ASASSN-15pz, while this feature is usually not seen in SNe Ia at such late phases.  \citet{Silverman2011} attributed such strong and long-lived \ion{Si}{2} $\lambda 6355$ absorption in SN 2009dc to synthesizing more silicon in the explosion than a typical SN Ia. At $1-2$ months, the spectra of SN 2012dn have line shapes and widths more similar to normal SNe Ia than SN 2009dc/ASASSN-15pz. The differences between SN 2012dn and SN 2009dc/ASASSN-15pz in the spectra at $1-2$ months are likely due to the higher ejecta velocities, as shown in Fig.~\ref{velocity}. The ejecta velocity of SN 2012dn derived from \ion{Si}{2} $\lambda 6355$ is $\sim 4000\,{\rm km\,s}^{-1}$ higher than SN 2009dc or $\sim 3000\,{\rm km\,s}^{-1}$ higher than ASASSN-15pz. \cite{Silverman2011} noted the interesting similarities in spectral features and line widths between SN 2009dc and SN 2002cx at these phases. One notable difference is the absence of \ion{Si}{2} $\lambda 6355$ absorption in SN 2002cx, while it is relatively strong for SN 2009dc/ASASSN-15pz. As we discuss in \S~\ref{nebular}, the later-phase spectra of ASASSN-15pz/SN 2009dc are substantially different from SN 2002cx.

\begin{figure*}
\centerline{\includegraphics[width=18cm]{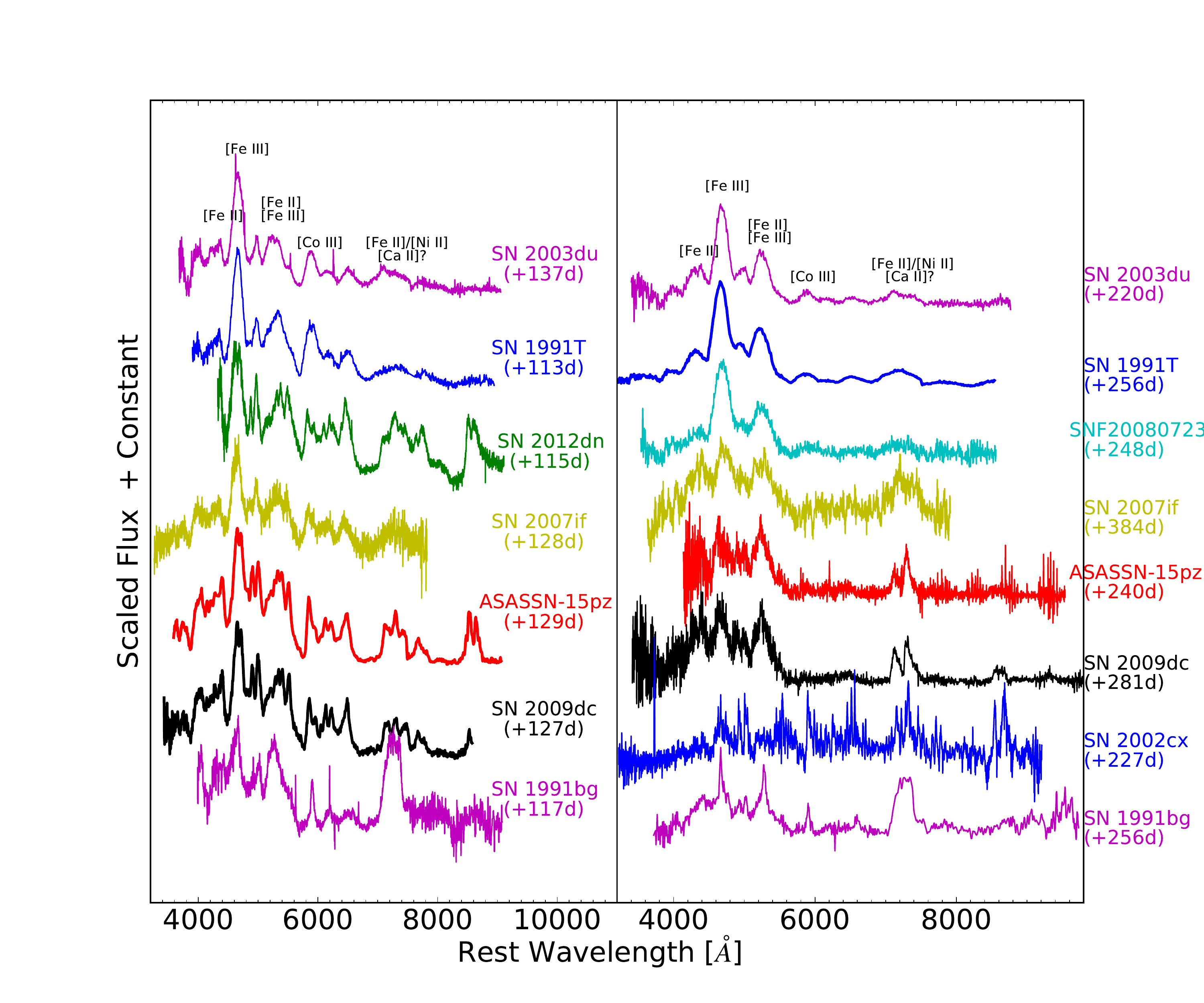}}
\caption{Left: early nebular-phase spectra of ASASSN-15pz compared with those of SNe Ia 2003du, SN 1991T, and SNe Ia-pec SN 2012dn \citep{Parrent2016}, SN 2007if \citep{Blondin2012}, and SN 2009dc. Right: nebular-phase spectra of ASASSN-15pz compared with those of SNe Ia 2003du \citep{Stanishev2007}, sub-luminous SN 1991bg \citep{Turatto1996}, over-luminous 
SN 1991T \citep{Gomez1998}, 1991T-like SN Ia SNF20080723 \citep{Scalzo2012}, and SNe Ia-pec SN 2007if \citep{Taubenberger2013}, SN 2009dc \citep{Silverman2011} as well as SN 2002cx \citep{Jha2006}.} 
\label{nebular_spectra}
\end{figure*}

\subsection{Nebular-phase spectra}
\label{nebular}
The supernova ejecta become (nearly) optically thin a few months after explosion. These Nebular-phase spectra probe the inner regions of the ejecta, often providing key insights into the explosion and clearly distinguishing different classes of supernovae. We show comparisons of the late-time spectra of ASASSN-15pz (+129 days and +240 days, respectively) with SNe Ia and Ia-like peculiar objects at similar phases in  Fig.~\ref{nebular_spectra}.

We label the prominent forbidden emission lines that shape the general nebular phase spectra of SNe Ia \citep[see, e.g.][]{Axelrod80, Maguire2018} in Fig. \ref{nebular_spectra}. The main \rm{Fe} and \rm{Co} features between $\sim 4000$~\AA\ and $\sim 8000$~\AA, are present in all sub-classes of SNe Ia but exhibit a broad range of line widths and velocity distributions \citep[see, e.g.,][]{Mazzali1998, Kushnir2013, Dong2015}.  The nebular spectra of ASASSN-15pz and SN 2009dc are well distinguished from normal SNe Ia. In spite of the spectroscopic differences from SN 2009dc at $1-2$ months, the late-time spectrum (+115 days) of SN~2012dn closely resembles those of SN 2009dc and ASASSN-15pz, providing good evidence that SN~2012dn is a member of the 2009dc-like SNe~Ia-pec class. Similarly, as shown in Fig.~\ref{nebular_spectra}, the nebular spectra of SN 2007if, especially the nebular spectrum at +384 days, is more similar to the 2009dc-like spectra than to those of the normal SNe~Ia.
 
At $\sim 4-5$ months (left panel of Fig.~\ref{nebular_spectra}), the differences between 2009dc-like SNe~Ia and SNe~Ia (SN 2003du and SN 1991T) are clearly seen in at least the following three aspects: (1) the nebular features of 2009dc-like SNe are generally sharper (i.e., lower velocities), and there are usually additional narrow features at the location of the main Fe or Co features of SNe Ia (e.g., the prominent [\ion{Fe}{2}]/[\ion{Fe}{3}] complex between 4400~\AA\ and 5700~\AA\ with two main broad peaks and also the [\ion{Co}{3}] region to their red);  (2) the feature around 7300\,\AA\, (likely including contributions from [\ion{Ca}{2}] $\lambda$7291, $\lambda$7324, and [\ion{Fe}{2}] $\lambda$7155) is relatively stronger than in SN~2003du and SN~1991T, and there is an additional emission feature at 7660\,\AA; (3) strong emission lines at around 8600\,\AA (Ca triplet) are seen in ASASSN-15pz and SN~2012dn. These are  are commonly seen in the nebular phase spectra of core-collapse supernovae, but not seen in SN~2003du and SN~1991T at similar phases \citep{Maguire2018}. Detailed modeling will be needed to understand whether the presence of strong \ion{Ca}{2} triplet nebular lines is compatible with models synthesizing a large of amount of $^{56}$Ni.

In  nebular-phase spectra obtained beyond 200 days after $B_{\rm max}$, 2009dc-like SNe~Ia, including ASASSN-15pz, are clearly distinguished from any other class of SNe Ia or Ia-like objects (from 1991bg-like SN Ia to 1991T-like SNe Ia also 2002cx-like SN~Ia-pec) as shown in the right panel of Fig.~\ref{nebular_spectra}. Thus, even if the very early-phase spectra are missing, it is possible to distinguish 2009dc-like SNe from SNe Ia and other Ia-like SNe. Similar to SNe~Ia, the blue part of the optical spectra of ASASSN-15pz and other 2009dc-like objects are dominated by [\ion{Fe}{2}]/[\ion{Fe}{3}] complexes between 4400\,\AA\, and 5700\,\AA\, but their shapes and line ratios are substantially different. As pointed out by \cite{Taubenberger2013}, the 5200\,\AA\, feature, a blend of [\ion{Fe}{2}] and [\ion{Fe}{3}], is stronger in 2009dc-like SNe~Ia than the features at 4700\,\AA\, that are primarily due to [\ion{Fe}{3}], implying a higher [\ion{Fe}{2}] fraction and a lower ionization state in the ejecta than normal SNe~Ia. The higher [\ion{Fe}{2}] to [\ion{Fe}{3}] ratio is also supported by the relatively strong feature at around 4300\,\AA, which is attributed to emission of [\ion{Fe}{2}] $\lambda$4244 and [\ion{Fe}{2}] $\lambda$4416 \citep{Mazzali2012, Mazzali2015}.  \citet{Taubenberger2013} suggested that the lower ionization states found in 2009dc-like SNe~Ia might be the result of enhanced recombination due to having higher inner ejecta densities due to the low expansion velocities of the 2009dc-like objects.

\section{Summary and Discussion}
\label{summary_discussion}

Our comprehensive photometric and spectroscopic data of ASASSN-15pz add to a handful of well-studied 2009dc-like objects. The main results of our analysis are as follow:

\begin{enumerate}
\item{ASASSN-15pz is a spectroscopic twin with the luminous SN Ia-pec SN~2009dc at all observed phases, and they share many common properties that distinguish them from the SNe~Ia population, such as being exceptionally UV bright at early phases, showing strong \ion{C}{2} $\lambda$6580 absorption at pre- and near-peak spectra, and having relatively slow expansion velocities.}
\item{With little host extinction, the peak luminosity of ASASSN-15pz is well determined  ($M_B^{\rm peak}=-19.69\pm0.12\,{\rm mag}$ and $M_V^{\rm peak}=-19.67\pm0.12\,{\rm mag}$), which is comparable with 1991T-like objects at the luminous end of SN Ia luminosity function but substantially dimmer than SN~2009dc.}
\item{From the bolometric light curve, the $^{56}$Ni mass synthesized in the explosion of ASASSN-15pz is $1.13 \pm 0.14\,M_\odot$, which is more than all normal SNe~Ia but lower than found for several 2009dc-like SNe Ia-pec. The $\gamma$-ray escape time scale of $t_0 = 46.4\pm0.5$ days is considerably higher than those of normal SNe~Ia, implying a relatively high ejecta column density.}
\item{ASASSN-15pz has broad $B$- and $V$-band light curves and its post-peak decline rate of $m_{15}(B) = 0.67\pm0.07\,{\rm mag}$ is similar to SN~2009dc and slower than all normal SNe~Ia. As a population, 2009dc-like SNe show a significant diversity in peak luminosities and do not follow the tight width-luminosity relation \citep{Phillips1993, Burns2014, Burns2018} of the SN Ia population. The NIR and late-time $\gtrsim50$\,days optical light curves of the three well-studied 2009dc-like objects (ASASSN-15pz, SN~2009dc and SN~2012dn) show a large diversity in their evolutions.}
\end{enumerate}

We have collected $\sim 200$ objects in our complete volume-limited $(z<0.02)$ sample of SNe~Ia and SNe~Ia-like objects (Chen, P. et al. 2019, in preparation). Of this sample ASASSN-15pz is the only 2009dc-like object, suggesting that the volumetric rate of 2009dc-like object is $\sim 1\%$ of the total SN Ia rate. In the 100 type IA Supernovae (100IAS) survey \citep{Dong2018}, we systematically collect nebular phase spectra from our volume-limited sample. Nebular-phase spectra can be crucial for understanding explosion physics by probing the inner ejecta. In particular, the nebular-phase spectra of ASASSN-15pz and other 2009dc-like objects (including SN 2007if, SN~2009dc, and SN~2012dn) are well separated from normal SNe~Ia, exhibiting characteristic features such as sharp nebular lines, low ionization states for Fe (low ratio of [\ion{Fe}{2}] emission to [\ion{Fe}{3}]), and strong [\ion{Ca}{2}] lines and \ion{Ca}{2} NIR triplet. Therefore, nebular spectra can provide definitive identification of 2009dc-like objects. The fact that 2009dc-like objects can exhibit significant   diversity in their photometric properties even between spectroscopic twins may provide an avenue to  constrain  the progenitors or explosion process leading to these events.  

The first few well-observed 2009dc-like SNe discovered since \citet{Howell2006} all reached high peak luminosities at $\sim -$20\,mag, leading to the speculation of super-Chandrasekhar-mass progenitors, although whether super-Chandrasekhar progenitors lead to 2009dc-like explosions is unknown 
(see, e.g., \citealt{Hachinger2012}, \citealt{Noebauer2016}, \citealt{Fink2018} and also the discussion in \citealt{Maoz2014}). 
The popular naming, ``Super-Chandrasekhar-mass SNe Ia,'' mainly rested upon the first 2009dc-like SNe being more luminous than all SNe Ia, whose origin were thought to be due to Chandrasekhar-mass explosions, according to popular theoretical notion. However, in recent years, there has been increasing observational and theoretical evidence against the Chandrasekhar-mass models as the main channel for the SN Ia population \citep[e.g., see reviews by][]{Maoz2014,Wang2018}. Here we find that 2009dc-like SNe Ia are not necessarily more luminous than all SNe Ia, which also undermines the empirical basis for the ``Super-Chandrasekhar'' naming. We therefore suggest referring to them in the future  as ``2009dc-like SNe~Ia-pec'' to avoid using a designation that is based on an uncertain theoretical interpretation.

\acknowledgments
We thank Doron Kushnir, Erika K. Carlon, Francesco Taddia, Barry Madore, Mark Seibert, Jeff Rich for their help. P.C., S.D., and S. B. acknowledge Project 11573003 supported by NSFC. This research uses data obtained through the Telescope Access Program (TAP), which has been funded by the National Astronomical Observatories of China, the Chinese Academy of Sciences, and the Special Fund for Astronomy from the Ministry of Finance.  K.M. acknowledges support from STFC (ST/M005348/1) and from H2020 through an ERC Starting Grant (758638). C.S.K, K.Z.S., and T.A.T. are supported by NSF grants AST-1515876, AST-1515927, and AST-1814440.  P.J.B. is supported by NASA through HST program \#14144 through a grant from the Space Telescope Science Institute, which is operated by the Association of Universities for Research in Astronomy, Inc., under NASA contract NAS 5-26555. Support for J.L.P. is provided in part by FONDECYT through the grant 1191038 and by the Ministry of Economy, Development, and Tourism's Millennium Science Initiative through grant IC120009, awarded to The Millennium Institute of Astrophysics, MAS.  This work is partly based on observations made with the Nordic Optical Telescope, operated by the Nordic Optical Telescope Scientific Association at the Observatorio del Roque de los Muchachos, La Palma, Spain, of the Instituto de Astrofisica de Canarias. The data presented here were obtained in part with ALFOSC, which is provided by the Instituto de Astrofisica de Andalucia (IAA) under a joint agreement with the University of Copenhagen and NOTSA. M.S. is supported in part by a generous grant (13261) from VILLUM FONDEN and a project grant from the Independent Research Fund Denmark. This work is based (in part) on observations collected at the European Organisation for Astronomical Research in the Southern Hemisphere, Chile, under ESO programme 0101.D-0202, and as part of PESSTO (the Public ESO Spectroscopic Survey for Transient Objects Survey) ESO program 188.D-3003, 191.D-0935, 197.D-1075.
We thank the Las Cumbres Observatory and its staff for its continuing support of the ASAS-SN project. ASAS-SN is supported by the Gordon and Betty Moore Foundation through grant GBMF5490 to the Ohio State University and NSF grant AST-1515927. Development of ASAS-SN has  been supported by NSF grant AST-0908816, the Mt. Cuba Astronomical Foundation, the Center for Cosmology and AstroParticle Physics at the Ohio State University, the Chinese Academy of Sciences South America Center for Astronomy (CAS- SACA), the Villum Foundation, and George Skestos. This paper uses data products produced by the OIR Telescope Data Center, supported by the Smithsonian Astrophysical Observatory. UKIRT is owned by the University of Hawaii (UH) and operated by the UH Institute for Astronomy; operations are enabled through the cooperation of the East Asian Observatory. When (some of) the data reported here were acquired, UKIRT was supported by NASA and operated under an agreement among the University of Hawaii, the University of Arizona, and Lockheed Martin Advanced Technology Center; operations were enabled through the cooperation of the East Asian Observatory.

 \software{Python, Astropy \citep{AstropyCollaboration2013}, Matplotlib \citep{Barrett2005}, ISIS image subtraction package \citep{Alard1998, Alard2000}, DoPHOT \citep{Schechter1993}, HEASOFT \citep{HEASARC2014}, IRAF \citep{Tody1986, Tody1993}}

\bibliographystyle{apj}
\bibliography{asassn-15pz}

\appendix

\section{Imaging Observations and Photometry}
\label{sec:photo}

ASAS-SN images are processed with an automated pipeline using the ISIS image subtraction package \citep{Alard1998, Alard2000} and calibrated with the AAVSO Photometric All-Sky Survey (APASS; \citealt{Henden2015}). 
Optical images of ASASSN-15pz are primarily from the Las Cumbres Observatory Global Telescope Network (LCOGT; \citealt{Brown2013}) 1 m telescopes at the Siding Spring Observatory (SSO), South African Astronomical Observatory (SAAO) and Cerro Tololo Interamerican Observatory (CTIO). 
We also obtained optical and simultaneous near-infrared (NIR) data with A Novel Dual Imaging CAMera (ANDICAM; \citealt{DePoy2003}) on the 1.3 m Small \& Moderate Aperture Research Telescope System (SMARTS; \citealt{Subasavage2010}) at Cerro Tololo, Chile.
In addition, we obtained NIR imaging observations with the NOTCam on Nordic Optical Telescope (NOT), the UKIRT Fast Track Imager (UFTI) and the Wide Field CAMera (WFCAM; \citealt{Casali2007}) on the United Kingdom Infrared Telescope (UKIRT). The optical and near-infrared images were reduced after bias/dark-frame and flat-field corrections. We use DoPHOT \citep{Schechter1993} to perform Point-spread-function (PSF) photometry, which is sufficient for ASASSN-15pz given the very low host-galaxy background. For the optical bands (Johnson $B$, $V$, and SDSS-$r$, $i$) the photometry data were calibrated using APASS, and the NIR photometry was calibrated using 2MASS \citep{Skrutskie2006}. We estimate the photometric uncertainties by adding the  DoPHOT and the zero-point calibration uncertainties in quadrature. The optical and NIR photometric results are given in Table~\ref{tab:LCOGT_BVri_mags_02} and \ref{tab:NIR_mags}, where the SDSS-$r$ and $i$ magnitudes are in the AB magnitude system and all the others are in the Vega system. 

Target of Opportunity (ToO) observations of ASASSN-15pz  (target ID 34102) were performed between Oct 14th 2015 and  Feb 27th 2016 with \textit{Swift}/UVOT\citep{Poole2008} in the $uvw2$ (1928\,\AA\,), $uvm2$ (2246 \,\AA\,), $uvw1$ (2600 \,\AA\,), $u$ (3465 \,\AA\,), $b$ (4392 \,\AA\,), and $v$ (5468 \,\AA\,) filters. Host-galaxy template images were obtained in March 2017 (PI M. Stritzinger, target ID 87296) to estimate and we then subtract the background galaxy contribution. We extracted source counts from a 5\farcs0 radius region centered on ASASSN-15pz using the UVOT software task \textit{uvotsource}. The source counts were converted into the Vega magnitude system based on the most recent UVOT calibrations \citep{Breeveld2011}, and the UVOT magnitudes are listed in Table~\ref{tab:Swift_mags}.

\begin{table*}
\scriptsize
\caption{LCOGT BVri magnitudes of ASASSN-15pz}
\begin{center}
\begin{tabular}{lcccccc}
\hline
\hline
Date & JD & phase$^a$ (day) & $B$ (mag) & $V$ (mag) & SDSS-$r$ (mag) & SDSS-$i$ (mag) \\
\hline
2015 Sep 28 & $2457294.2712$ & $-12.95$ & $15.61\pm0.09$ & $15.49\pm0.05$ & $15.59\pm0.05$ & $15.77\pm0.07$ \\
 2015 Sep 30 & $2457295.6212$ & $-11.60$ & $15.21\pm0.04$ & $15.15\pm0.03$ & $15.26\pm0.04$ & $15.54\pm0.03$ \\
 2015 Oct 1 & $2457297.0102$ & $-10.21$ & $14.99\pm0.03$ & $14.91\pm0.04$ & $15.04\pm0.03$ & $15.32\pm0.05$ \\
 2015 Oct 2 & $2457298.4970$ & $-8.72$  & $14.73\pm0.05$ & $14.66\pm0.07$ & $14.87\pm0.10$ & $15.12\pm0.08$ \\
 2015 Oct 4 & $2457300.3482$ & $-6.87$  & $14.52\pm0.03$ & $14.51\pm0.02$ & $14.64\pm0.04$ &        ...                \\
 2015 Oct 6 & $2457302.3425$ & $-4.88$  & $14.36\pm0.02$ & $14.36\pm0.02$ & $14.51\pm0.05$ & $14.95\pm0.03$ \\
 2015 Oct 8 & $2457304.3372$ & $-2.88$  & $14.28\pm0.02$ & $14.28\pm0.03$ & $14.41\pm0.06$ & $14.87\pm0.04$ \\
 2015 Oct 10 & $2457306.3317$ & $-0.89$  & $14.23\pm0.02$ & $14.23\pm0.02$ & $14.37\pm0.03$ & $14.86\pm0.03$ \\
 2015 Oct 12 & $2457308.3824$ & $1.16$   & $14.19\pm0.05$ & $14.23\pm0.03$ & $14.36\pm0.03$ & $14.87\pm0.04$ \\
 2015 Oct 15 & $2457310.9940$ & $3.77$   & $14.32\pm0.04$ & $14.28\pm0.02$ & $14.38\pm0.03$ & $14.87\pm0.05$ \\
 2015 Oct 17 & $2457313.1149$ & $5.89$   & $14.34\pm0.04$ & $14.34\pm0.08$ & $14.41\pm0.03$ & $14.91\pm0.07$ \\
 2015 Oct 20 & $2457316.3547$ & $9.13$   & $14.50\pm0.02$ & $14.40\pm0.03$ &       ...                 & $14.95\pm0.03$ \\
 2015 Oct 24 & $2457319.5479$ & $12.33$  & $14.68\pm0.03$ & $14.51\pm0.03$ & $14.60\pm0.04$ & $15.04\pm0.05$ \\
 2015 Oct 26 & $2457322.3168$ & $15.10$  & $14.91\pm0.03$ &         ...               & $14.72\pm0.04$ & $15.16\pm0.04$ \\
 2015 Oct 29 & $2457325.2811$ & $18.06$  & $15.12\pm0.04$ & $14.73\pm0.03$ & $14.79\pm0.04$ & $15.16\pm0.04$ \\
 2015 Nov 1 & $2457327.7639$ & $20.54$  & $15.30\pm0.03$ & $14.84\pm0.05$ & $14.87\pm0.04$ & $15.15\pm0.04$ \\
 2015 Nov 3 & $2457330.2885$ & $23.07$  & $15.47\pm0.01$ & $14.93\pm0.03$ & $14.89\pm0.03$ & $15.13\pm0.04$ \\
 2015 Nov 6 & $2457333.0766$ & $25.86$  & $15.69\pm0.05$ & $15.05\pm0.04$ & $14.96\pm0.03$ & $15.16\pm0.04$ \\
 2015 Nov 9 & $2457336.0004$ & $28.78$  & $15.92\pm0.06$ & $15.17\pm0.04$ & $15.06\pm0.09$ & $15.13\pm0.08$ \\
 2015 Nov 15 & $2457342.1386$ & $34.92$  & $16.36\pm0.04$ & $15.52\pm0.03$ & $15.31\pm0.04$ & $15.32\pm0.04$ \\
 2015 Nov 18 & $2457345.0433$ & $37.82$  &           ...              & $15.66\pm0.03$ & $15.45\pm0.04$ & $15.45\pm0.05$ \\
 2015 Nov 19 & $2457345.5801$ & $38.36$  & $16.61\pm0.05$ & $15.68\pm0.03$ & $15.45\pm0.03$ & $15.47\pm0.04$ \\
 2015 Nov 21 & $2457348.0275$ & $40.81$  &         ...               & $15.79\pm0.03$ & $15.57\pm0.04$ & $15.62\pm0.04$ \\
 2015 Nov 24 & $2457351.0198$ & $43.80$  & $16.74\pm0.06$ & $15.89\pm0.04$ & $15.73\pm0.04$ & $15.77\pm0.05$ \\
 2015 Nov 27 & $2457354.0115$ & $46.79$  &        ...                 &       ...                  & $15.84\pm0.06$ &     ...                   \\  
 2015 Nov 29 & $2457356.0039$ & $48.78$  & $16.91\pm0.04$ & $16.00\pm0.04$ & $15.87\pm0.03$ & $15.93\pm0.04$ \\
 2015 Dec 1 & $2457358.4187$ & $51.20$  & $16.92\pm0.06$ & $16.09\pm0.04$ & $15.97\pm0.04$ & $16.03\pm0.05$ \\
 2015 Dec 4 & $2457361.2791$ & $54.06$  & $17.01\pm0.04$ & $16.18\pm0.03$ & $16.08\pm0.03$ & $16.23\pm0.05$ \\
 2015 Dec 6 & $2457363.2821$ & $56.06$  & $17.05\pm0.05$ & $16.22\pm0.05$ & $16.12\pm0.04$ & $16.23\pm0.05$ \\
 2015 Dec 8 & $2457365.3853$ & $58.17$  & $17.15\pm0.05$ & $16.28\pm0.04$ & $16.18\pm0.04$ & $16.32\pm0.06$ \\
 2015 Dec 11 & $2457368.3814$ & $61.16$  & $17.20\pm0.05$ & $16.36\pm0.03$ & $16.32\pm0.04$ & $16.44\pm0.04$ \\
 2015 Dec 16 & $2457372.6042$ & $65.38$  & $17.24\pm0.07$ & $16.44\pm0.05$ & $16.44\pm0.04$ & $16.57\pm0.06$ \\
 2015 Dec 18 & $2457375.0004$ & $67.78$  & $17.31\pm0.10$ & $16.53\pm0.07$ & $16.43\pm0.09$ & $16.62\pm0.07$ \\
 2015 Dec 21 & $2457378.2918$ & $71.07$  & $17.38\pm0.11$ & $16.68\pm0.07$ & $16.53\pm0.09$ & $16.87\pm0.10$ \\
 2015 Dec 24 & $2457380.9518$ & $73.73$  & $17.52\pm0.10$ & $16.70\pm0.07$ & $16.68\pm0.06$ &       ...                 \\  
 2015 Dec 26 & $2457383.3157$ & $76.10$  & $17.55\pm0.06$ & $16.73\pm0.04$ & $16.77\pm0.04$ & $16.98\pm0.07$ \\
 2015 Dec 29 & $2457386.1151$ & $78.90$  & $17.56\pm0.08$ &      ...                   &         ...               &        ...                 \\  
 2015 Dec 30 & $2457387.1286$ & $79.91$  &        ...                &       ...                   &         ...               & $17.07\pm0.10$ \\
 2016 Jan 1 & $2457389.3087$ & $82.09$  & $17.64\pm0.06$ &       ...                  &         ...               & $17.22\pm0.07$ \\
 2016 Jan 2 & $2457390.3237$ & $83.10$  & $17.68\pm0.08$ & $16.97\pm0.04$ & $17.00\pm0.04$ & $17.25\pm0.07$ \\
 2016 Jan 4 & $2457392.4078$ & $85.19$  & $17.71\pm0.10$ & $17.02\pm0.07$ &                & $17.24\pm0.09$ \\
 2016 Jan 7 & $2457395.3091$ & $88.09$  & $17.77\pm0.07$ & $17.04\pm0.05$ & $17.13\pm0.07$ & $17.28\pm0.10$ \\
 2016 Jan 9 & $2457396.6167$ & $89.40$  & $17.77\pm0.07$ & $17.11\pm0.06$ & $17.19\pm0.09$ & $17.40\pm0.08$ \\
 2016 Jan 10 & $2457398.3758$ & $91.16$  & $17.88\pm0.06$ & $17.18\pm0.04$ & $17.30\pm0.05$ & $17.56\pm0.10$ \\
 2016 Jan 17 & $2457404.6003$ & $97.38$  & $18.01\pm0.09$ & $17.37\pm0.06$ & $17.53\pm0.08$ & $17.56\pm0.09$ \\
 2016 Jan 23 & $2457411.0425$ & $103.82$ & $18.11\pm0.14$ & $17.56\pm0.08$ & $17.67\pm0.08$ & $17.91\pm0.14$ \\
 2016 Jan 27 & $2457415.2879$ & $108.07$ & $18.20\pm0.07$ & $17.54\pm0.06$ & $17.88\pm0.05$ & $18.01\pm0.09$ \\
 2016 Feb 6 & $2457425.3006$ & $118.08$ & $18.38\pm0.16$ &        ...                 & $18.27\pm0.10$ & $18.20\pm0.23$ \\
 2016 Feb 11 & $2457430.3590$ & $123.14$ & $18.49\pm0.03$ & $18.01\pm0.09$ & $18.42\pm0.09$ & $18.39\pm0.16$ \\
 2016 Feb 18 & $2457437.3258$ & $130.11$ & $18.63\pm0.06$ & $18.10\pm0.04$ & $18.47\pm0.06$ & $18.63\pm0.08$ \\
 2016 Feb 22 & $2457441.2663$ & $134.05$ & $18.74\pm0.12$ & $18.18\pm0.06$ & $18.80\pm0.15$ & $18.68\pm0.10$ \\
 2016 Mar 1 & $2457449.2720$ & $142.05$ & $18.86\pm0.06$ & $18.35\pm0.06$ & $18.91\pm0.06$ & $18.94\pm0.10$ \\
 2016 Mar 6 & $2457453.5411$ & $146.32$ &       ...                  & $18.53\pm0.06$ &      ...                   & $19.25\pm0.16$ \\
 2016 Mar 12 & $2457460.2677$ & $153.05$ & $19.04\pm0.08$ & $18.67\pm0.07$ & $19.44\pm0.11$ & $19.21\pm0.15$ \\
 2016 Mar 17 & $2457465.2448$ & $158.02$ & $19.11\pm0.08$ & $18.72\pm0.07$ & $19.43\pm0.15$ & $19.40\pm0.15$ \\
 2016 Mar 22 & $2457470.2461$ & $163.03$ &      ...                  & $18.95\pm0.15$ &     ...                     & $19.57\pm0.29$ \\
 \hline
\end{tabular}
\end{center}
$^a$ relative to B-band peak time at JD=2457307.2.
\label{tab:LCOGT_BVri_mags_02}
\end{table*}

\begin{table*}
\scriptsize
\caption{SMARTS, NOT and UKIRT magnitudes of ASASSN-15pz}
\begin{center}
\begin{tabular}{lcccccccc}
\hline
\hline
Date & JD & phase$^a$ (day) & $B$ (mag) & $V$ (mag) &$J$ (mag) & $H$ (mag) & $K$ (mag)& Instrument \\
\hline
 2015 Oct 23 & $2457318.7216$ & $11.50$  & $14.74\pm0.07$ & $14.50\pm0.07$ &$15.11\pm0.05$ & $14.87\pm0.05$ &    ...         & ANDICAM$^b$ \\
 2015 Oct 27 & $2457322.6689$ & $15.45$  & $15.02\pm0.07$ & $14.63\pm0.07$ &$15.39\pm0.05$ & $14.72\pm0.05$ &    ...         & ANDICAM \\
 2015 Oct 28 & $2457323.5641$ & $16.34$  &   ...          &         ...    &$15.28\pm0.06$ & $14.79\pm0.12$ & $14.67\pm0.09$ & NOTCam$^c$  \\  
 2015 Oct 30 & $2457325.9293$ & $18.71$  &   ...          &         ...    &$15.36\pm0.11$ & $14.73\pm0.15$ & $14.78\pm0.20$ & UFTI$^d$    \\  
 2015 Nov 1 & $2457327.6091$ & $20.39$  & $15.40\pm0.07$ & $14.83\pm0.07$ &$15.51\pm0.06$ & $14.79\pm0.06$ &     ...        & ANDICAM \\
 2015 Nov 3 & $2457329.6998$ & $22.48$  & $15.58\pm0.06$ & $14.93\pm0.06$ &$15.44\pm0.05$ & $14.79\pm0.05$ &     ...        & ANDICAM \\
 2015 Nov 6 & $2457332.6043$ & $25.38$  & $15.79\pm0.07$ & $15.04\pm0.07$ &$15.51\pm0.06$ & $14.74\pm0.05$ &      ...       & ANDICAM \\
 2015 Nov 8 & $2457334.7238$ & $27.50$  & $15.95\pm0.07$ & $15.13\pm0.07$ &$15.51\pm0.05$ & $14.72\pm0.05$ &      ...       & ANDICAM \\
 2015 Nov 8 & $2457334.9057$ & $27.69$  &   ...          &         ...    &$15.40\pm0.10$ & $14.55\pm0.15$ & $14.84\pm0.20$ & UFTI    \\  
 2015 Nov 10 & $2457336.6209$ & $29.40$  & $16.09\pm0.06$ & $15.24\pm0.06$ &     ...       & $14.70\pm0.06$ &        ...     & ANDICAM \\
 2015 Nov 10 & $2457336.9105$ & $29.69$  &   ...          &         ...    &$15.39\pm0.10$ & $14.62\pm0.15$ & $14.87\pm0.20$ & UFTI    \\  
 2015 Nov 14 & $2457340.7524$ & $33.53$  & $16.45\pm0.07$ & $15.46\pm0.07$ &$15.47\pm0.05$ & $14.71\pm0.06$ &      ...       & ANDICAM \\
 2015 Nov 16 & $2457342.6969$ & $35.48$  & $16.61\pm0.07$ & $15.59\pm0.07$ &     ...       &      ...       &      ...       & ANDICAM \\  
 2015 Nov 20 & $2457346.7306$ & $39.51$  & $16.76\pm0.07$ & $15.76\pm0.07$ &     ...       &      ...       &      ...       &ANDICAM \\  
 2015 Nov 21 & $2457347.5969$ & $40.38$  & $16.81\pm0.07$ & $15.78\pm0.07$ &$15.57\pm0.06$ &        ...     &        ...     & ANDICAM \\
 2015 Nov 23 & $2457349.7622$ & $42.54$  & $16.86\pm0.07$ & $15.85\pm0.07$ &$15.77\pm0.06$ & $14.89\pm0.06$ &                & ANDICAM \\
 2015 Nov 24 & $2457350.5149$ & $43.29$  &   ...          &         ...    &$15.81\pm0.07$ & $15.07\pm0.10$ & $15.07\pm0.15$ & NOTCam  \\  
 2015 Nov 27 & $2457353.7222$ & $46.50$  & $16.98\pm0.07$ & $15.98\pm0.07$ &     ...       &      ...       &      ...       & ANDICAM \\  
 2015 Nov 29 & $2457355.6936$ & $48.47$  & $17.02\pm0.07$ & $16.03\pm0.07$ &$16.14\pm0.06$ &       ...      &          ...      & ANDICAM \\
 2015 Nov 30 & $2457356.6619$ & $49.44$  & $17.03\pm0.06$ & $16.06\pm0.06$ &$16.17\pm0.06$ & $15.10\pm0.06$ &       ...         & ANDICAM \\
 2015 Dec 3 & $2457359.7406$ & $52.52$  & $17.07\pm0.08$ & $16.14\pm0.08$ &     ...       &      ...       &      ...       & ANDICAM \\  
 2015 Dec 5 & $2457361.6426$ & $54.42$  & $17.13\pm0.06$ & $16.20\pm0.06$ &$16.37\pm0.06$ & $15.31\pm0.06$ &       ...         & ANDICAM \\
 2015 Dec 6 & $2457362.6341$ & $55.41$  & $17.13\pm0.06$ & $16.21\pm0.06$ &$16.46\pm0.06$ & $15.38\pm0.06$ &       ...         & ANDICAM \\
 2015 Dec 8 & $2457364.6270$ & $57.41$  & $17.17\pm0.07$ & $16.27\pm0.07$ &$16.62\pm0.07$ & $15.38\pm0.06$ &       ...         & ANDICAM \\
 2015 Dec 11 & $2457367.6116$ & $60.39$  & $17.26\pm0.07$ & $16.35\pm0.07$ &$16.68\pm0.07$ &      ...       &        ...        & ANDICAM \\
 2015 Dec 15 & $2457371.6651$ & $64.45$  & $17.31\pm0.07$ & $16.45\pm0.07$ &$16.90\pm0.07$ & $15.54\pm0.07$ &       ...         & ANDICAM \\
 2015 Dec 17 & $2457373.5966$ & $66.38$  & $17.36\pm0.07$ & $16.52\pm0.07$ &$17.16\pm0.09$ & $15.61\pm0.07$ &       ...         & ANDICAM \\
 2015 Dec 21 & $2457377.5802$ & $70.36$  & $17.45\pm0.07$ & $16.61\pm0.07$ &$17.24\pm0.09$ & $15.82\pm0.07$ &       ...         & ANDICAM \\
 2015 Dec 22 & $2457379.4459$ & $72.23$  &   ...          &         ...    &        ...    &        ...     & $15.36\pm0.17$   & NOTCam  \\  
 2015 Dec 23 & $2457379.5940$ & $72.37$  & $17.47\pm0.07$ & $16.65\pm0.07$ &     ...       &      ...       &      ...       & ANDICAM \\
 2015 Dec 24 & $2457380.7604$ & $73.54$  &   ...          &         ...    &$17.25\pm0.05$ & $15.90\pm0.08$ & $15.35\pm0.07$   & WFCAM$^e$   \\  
 2015 Dec 28 & $2457384.6249$ & $77.40$  & $17.59\pm0.07$ & $16.78\pm0.07$ &$17.57\pm0.11$ &      ...       &        ...       & ANDICAM \\
 2015 Dec 30 & $2457386.6397$ & $79.42$  & $17.63\pm0.07$ & $16.84\pm0.07$ &     ...       &      ...       &      ...       & ANDICAM \\
 2016 Jan 1 & $2457388.5718$ & $81.35$  & $17.67\pm0.07$ & $16.87\pm0.07$ &     ...       &      ...       &      ...       & ANDICAM \\
 2016 Jan 1 & $2457388.7285$ & $81.51$  &   ...          &         ...    &$17.59\pm0.05$ & $16.14\pm0.08$ & $15.46\pm0.07$   & WFCAM   \\  
 2016 Jan 3 & $2457390.5613$ & $83.34$  & $17.67\pm0.07$ & $16.92\pm0.07$ &     ...       &      ...       &      ...       & ANDICAM \\
 2016 Jan 3 & $2457390.7389$ & $83.52$  &   ...          &         ...    &$17.67\pm0.05$ & $16.22\pm0.08$ & $15.50\pm0.07$   & WFCAM   \\  
 2016 Jan 7 & $2457394.6230$ & $87.40$  & $17.80\pm0.07$ & $17.07\pm0.07$ &     ...       &      ...       &      ...       & ANDICAM \\
 2016 Jan 9 & $2457396.5623$ & $89.34$  & $17.81\pm0.07$ & $17.09\pm0.07$ &     ...       &      ...       &      ...       & ANDICAM \\
 2016 Jan 11 & $2457398.5604$ & $91.34$  & $17.88\pm0.07$ & $17.15\pm0.07$ &     ...       &      ...       &      ...       & ANDICAM \\
 2016 Jan 13 & $2457400.5831$ & $93.36$  & $17.92\pm0.07$ & $17.21\pm0.07$ &     ...       &      ...       &      ...       & ANDICAM \\
 2016 Jan 15 & $2457402.5868$ & $95.37$  & $17.98\pm0.07$ & $17.25\pm0.07$ &     ...       &      ...       &      ...       & ANDICAM \\
 2016 Jan 17 & $2457404.5331$ & $97.31$  & $18.02\pm0.07$ & $17.32\pm0.07$ &     ...       &      ...       &      ...       & ANDICAM \\
 2016 Jan 18 & $2457405.7279$ & $98.51$  &   ...          &         ...    &$18.39\pm0.06$ & $16.77\pm0.08$ & $15.88\pm0.09$   & WFCAM   \\  
 2016 Jan 19 & $2457406.5906$ & $99.37$  & $18.05\pm0.07$ & $99.90\pm0.07$ &     ...       &      ...       &      ...       & ANDICAM \\
 2016 Jan 21 & $2457408.5669$ & $101.35$ & $18.10\pm0.06$ & $17.42\pm0.06$ &     ...       &      ...       &      ...       & ANDICAM \\
 2016 Jan 24 & $2457412.3189$ & $105.10$ &   ...          &         ...    &$18.57\pm0.08$ & $16.92\pm0.15$ &      ...         & NOTCam  \\  
 2016 Jan 26 & $2457413.6057$ & $106.39$ & $18.22\pm0.10$ & $17.56\pm0.10$ &     ...       &      ...       &      ...       & ANDICAM \\\hline
\end{tabular}
\end{center}
$^a$ Relative to $B$-band peak time at JD=2457307.2; \\
$^b$ Optical and NIR camera on the SMARTS 1.3 m telescope; \\
$^c$ Near-infrared Camera and spectrograph on the Nordic Optical Telescope (NOT; 2.56 m);\\
$^d$ NIR imaging camera on the United Kingdom Infrared Telescope (UKIRT; 3.8 m);\\
$^e$ NIR imaging camera on the United Kingdom Infrared Telescope (UKIRT; 3.8 m);
\label{tab:NIR_mags}
\end{table*}

\begin{table*}
%\small
\scriptsize
\caption{ \textit{Swift} UVOT photometry of ASASSN-15pz}
\begin{center}
\begin{tabular}{lccccccc}
\hline
Date & JD & $uvw2$ & $uvm2$ & $uvw1$ & $u$ & $b$ & $v$ \\
\hline
 2015 Oct 14 & $2457309.8480$ & $15.75\pm0.05$ & $15.60\pm0.04$ & $14.60\pm0.04$ & $13.45\pm0.04$ & $14.24\pm0.03$ & $14.24\pm0.04$ \\
 2015 Oct 17 & $2457312.5954$ & $15.95\pm0.05$ & $15.94\pm0.05$ & $14.89\pm0.04$ & $13.66\pm0.04$ & $14.30\pm0.03$ & $14.35\pm0.04$ \\
 2015 Oct 20 & $2457315.7142$ & $16.36\pm0.05$ & $16.39\pm0.05$ & $15.25\pm0.05$ & $13.97\pm0.04$ & $14.46\pm0.03$ & $14.41\pm0.04$ \\
 2015 Oct 23 & $2457318.7697$ & $16.85\pm0.06$ & $16.85\pm0.06$ & $15.65\pm0.05$ & $14.37\pm0.04$ & $14.65\pm0.03$ & $14.50\pm0.04$ \\
 2015 Oct 26 & $2457321.7614$ & $17.09\pm0.07$ & $17.05\pm0.06$ & $15.96\pm0.05$ & $14.69\pm0.04$ & $14.84\pm0.03$ & $14.68\pm0.04$ \\
 2015 Oct 29 & $2457324.6257$ & $17.15\pm0.07$ & $17.16\pm0.07$ & $16.32\pm0.06$ & $14.99\pm0.04$ & $15.01\pm0.04$ & $14.68\pm0.04$ \\
 2015 Nov 1 & $2457327.6927$ & $17.51\pm0.12$ &     ...        & $16.50\pm0.07$ & $15.30\pm0.05$ & $15.27\pm0.04$ &      ...       \\
 2015 Nov 4 & $2457330.5998$ & $17.98\pm0.09$ & $17.66\pm0.08$ & $16.70\pm0.07$ & $15.56\pm0.05$ & $15.50\pm0.04$ & $14.94\pm0.04$ \\
 2015 Nov 7 & $2457333.7948$ & $17.89\pm0.09$ & $17.55\pm0.08$ & $16.86\pm0.07$ & $15.81\pm0.05$ & $15.74\pm0.04$ & $15.09\pm0.04$ \\
 2015 Nov 17 & $2457344.3791$ & $18.23\pm0.12$ & $17.86\pm0.10$ & $17.48\pm0.10$ & $16.60\pm0.08$ & $16.54\pm0.06$ & $15.60\pm0.06$ \\
 2015 Nov 22 & $2457349.2347$ & $18.39\pm0.15$ & $17.86\pm0.11$ & $17.75\pm0.12$ & $16.90\pm0.10$ & $16.66\pm0.07$ & $15.85\pm0.07$ \\
 2015 Nov 27 & $2457354.2926$ &      ...       &       ...      & $17.83\pm0.12$ & $17.02\pm0.10$ &     ...        &     ...        \\
 2015 Dec 2 & $2457359.0546$ & $18.72\pm0.20$ & $18.51\pm0.17$ & $17.88\pm0.15$ & $17.12\pm0.12$ & $16.93\pm0.08$ & $16.19\pm0.09$ \\
 2015 Dec 7 & $2457363.7879$ & $18.88\pm0.17$ & $18.46\pm0.13$ & $17.94\pm0.12$ & $17.29\pm0.10$ & $17.05\pm0.07$ & $16.22\pm0.07$ \\
 2015 Dec 12 & $2457369.3009$ & $19.16\pm0.22$ & $18.92\pm0.19$ & $17.94\pm0.13$ & $17.48\pm0.12$ & $17.24\pm0.08$ & $16.45\pm0.09$ \\
 2015 Dec 17 & $2457373.6811$ & $18.82\pm0.20$ & $18.88\pm0.21$ & $18.02\pm0.17$ & $17.53\pm0.15$ & $17.27\pm0.09$ & $16.66\pm0.12$ \\
 2015 Dec 22 & $2457379.1334$ & $19.21\pm0.20$ & $18.79\pm0.14$ & $18.19\pm0.14$ & $17.53\pm0.11$ & $17.41\pm0.07$ & $16.67\pm0.09$ \\
 2015 Dec 27 & $2457383.9241$ & $19.17\pm0.20$ & $19.00\pm0.17$ & $18.43\pm0.16$ & $17.77\pm0.13$ & $17.50\pm0.08$ & $16.76\pm0.10$ \\
 2016 Jan 1 & $2457388.8610$ & $19.40\pm0.23$ & $19.08\pm0.19$ & $18.84\pm0.21$ & $17.93\pm0.15$ & $17.40\pm0.08$ & $17.12\pm0.12$ \\
 2016 Jan 6 & $2457394.2360$ & $19.36\pm0.22$ & $19.00\pm0.17$ & $18.83\pm0.20$ & $17.95\pm0.14$ & $17.73\pm0.09$ & $17.21\pm0.12$ \\
 2016 Jan 10 & $2457398.2922$ & $19.63\pm0.26$ & $19.65\pm0.26$ & $18.80\pm0.20$ & $18.47\pm0.22$ & $17.81\pm0.10$ & $17.22\pm0.12$ \\
 2016 Jan 28 & $2457415.6343$ &  ...           & $19.63\pm0.25$ & $19.39\pm0.29$ & $18.84\pm0.26$ & $18.02\pm0.11$ & $17.68\pm0.16$ \\
 2016 Feb 7 & $2457425.6012$ &  ...           & ...            & ...             & $18.81\pm0.45$ & $18.32\pm0.23$ & $17.63\pm0.24$ \\
 2016 Feb 17 & $2457435.5204$ &  ...           & ...            & ...             & $19.00\pm0.32$ & $18.83\pm0.20$ & $17.95\pm0.21$ \\
 2016 Feb 27 & $2457446.3459$ &  ...           & ...            & ...             & $19.66\pm0.57$ & $18.83\pm0.23$ & $18.02\pm0.24$ \\
\hline
\end{tabular}
\end{center}
\label{tab:Swift_mags}
\end{table*}

\section{Spectroscopic Observations}
\label{sec:spectroscopy}

Spectroscopic observations were made at eight different epochs, ranging in phase from $-11.6$ days (JD 2457295.66) to +239.7 days (JD 2457546.90), as summarized in Table~\ref{spectroscopic_log}.  We made use of multiple instruments, including the 2.56 m Nordic Optical Telescope (NOT) with ALFOSC, the 2.5 m du Pont telescope with WFCCD, the 3.58 m New Technology Telescope (NTT) with EFOSC, and the 6.5 m \textit{Magellan} telescopes with IMACS and LDSS3C.
 The NTT spectra were obtained as part of the Public ESO Spectroscopic Survey for Transient Objects (PESSTO, \citealt{Smartt2015}) program and were reduced using a custom pipeline as described in \cite{Smartt2015}. The nebular-phase spectra taken at 151 days and 240 days with the \textit{Magellan} telescopes are part of the 100IAS program \citep{Dong2018}.  Spectroscopic data were reduced within the \texttt{IRAF} environment following  standard procedures. Appropriate spectrophotometric standard stars were used to compute nightly response functions, which were used to  flux-calibrate the science spectra.

\begin{table*}
\caption{Log of spectroscopic observation of ASASSN-15pz}
\begin{center}
\begin{tabular}{lcccc}
\hline
\hline
 Date    &   JD   &  Phase (d) &    Range(\AA)     &      Instrument  \\
 \hline
2015 Sep 30&2457295.66&-11.6&3500-9000&NOT$^a$/ALFOSC\\
2015 Nov 5 &2457331.54&+24.3&3500-9000&NOT/ALFOSC\\
2015 Nov 9&2457335.71&+28.5&3700-9250&du Pont$^b$/WFCCD\\
2015 Dec 17&2457373.74&+66.5&3650-9250&NTT$^c$/EFOSC\\
2016 Jan 11&2457398.57&+91.3&3700-9250&du Pont/WFCCD\\
2016 Feb 2&2457420.56&+113.3&3700-9250&du Pont/WFCCD\\
2016 Feb 18&2457436.56&+129.3&3650-9250&NTT/EFOSC\\
2016 Feb 18&2457436.60&+129.4&3650-9250&NTT/EFOSC\\
2016 Mar 11&2457458.54&+151.3&4200-9400&Baade$^d$/IMACS\\
2016 Jun 7&2457546.90&+239.7&4800-10000&Clay$^e$/LDSS3C\\
\hline 
\end{tabular}
\label{tab:speclog}
\end{center}
$^a$ Nordic Optical Telescope (2.56 m) located at Roque de los Muchachos Observatory, La Palma in the Canary Islands;
$^b$ du Pont telescope (2.5 m) located in Chile at Las Campanas Observatory;
$^c$ New Technology Telescope (3.58 m) located in Chile at the La Silla Observatory; 
$^d$ One of the Magellan Telescopes (6.5 m) located at Las Campanas Observatory in Chile;
$^e$ One of the Magellan Telescopes (6.5 m) located at Las Campanas Observatory in Chile
\label{spectroscopic_log}
\end{table*}

\end{document}